\documentclass[prd,showpacs,superscriptaddress,nofootinbib,floatfix,showkeys,11pt]{revtex4}
\usepackage{graphicx}
\usepackage{amsmath}
\usepackage{bm}
\usepackage{yhmath}
\usepackage{mathtools}
\usepackage{wasysym}
\usepackage[colorlinks,citecolor=violet,urlcolor=violet,linkcolor=blue]{hyperref}
\usepackage{subfigure}
\usepackage{color}
\usepackage{cases}
\usepackage{subfigure}
\usepackage{times}
\usepackage{dcolumn,booktabs,bm}
\usepackage{slashed}
\usepackage{amsfonts,amssymb,stmaryrd,latexsym,amsmath}
\usepackage{textcomp}
\usepackage{multirow}
\usepackage{cancel}
\usepackage{array}
\usepackage{orcidlink}
\usepackage{longtable}

\def\SU3{{\text{SU(3)}_{\rm F}}}

\def \pcs4338{{P_{\psi s}^\Lambda(4338)^0}}

\allowdisplaybreaks

\begin{document}
	
	\title{\textcolor{violet}{Reissner-Nordström Black Holes in Quintic Quasi-topological Gravity}}
	
	\author{A. R. Olamaei\,\orcidlink{0000-0003-3529-3002}}\email{olamaei@jahromu.ac.ir}\thanks{Corresponding Author}
	\affiliation{Department of Physics, Jahrom University, Jahrom, P.~ O.~ Box 74137-66171, Iran}
	
	\author{A.~Bazrafshan\,}
	\affiliation{Department of Physics, Jahrom University, Jahrom, P.~ O.~ Box 74137-66171, Iran}
	
	\author{M. Ghanaatian\,\orcidlink{0000-0003-3741-2167}}
	\affiliation{Department of Physics, Jahrom University, Jahrom, P.~ O.~ Box 74137-66171, Iran}
	
	
	\begin{abstract}
		This paper investigates charged black holes within the framework of quintic quasi-topological gravity, focusing on their thermodynamics, conserved quantities, and stability. We construct numerical solutions and explore their thermodynamic properties, supplemented by the study of analytically solvable special cases. By verifying the first law of thermodynamics, we validate our approach and compare our findings to those of Einstein gravity. The physical properties of the solutions are examined across anti-de Sitter, de Sitter, and flat spacetime backgrounds. Our analysis reveals that anti-de Sitter solutions demonstrate thermal stability, while de Sitter and flat solutions lack this property. Finally, we discuss the implications of our results and propose potential avenues for future research in this field.

	\end{abstract}
	
	\maketitle
	
	\thispagestyle{empty}
	
	\textcolor{violet}{\section{Introduction}\label{Int}}
	
	The AdS/CFT correspondence, also known as the gauge/gravity duality \cite{Maldacena:1997re}, represents a remarkable phenomenon whereby supergravity in a certain dimensionality is dually related to a conformal field theory residing in a lower dimension. This profound duality bestows upon us the miraculous ability to compute a particular quantity on one side of the correspondence (referred to as the gravity side), which may prove challenging or even infeasible to calculate on the other side (referred to as the gauge side). Such an extraordinary feature opens up new avenues for investigation and insights into the intricate interplay between gravity and quantum field theory.
	
	In the context of Einstein gravity governed by a two-derivative bulk action, the applicability of the AdS/CFT conjecture is restricted to the regime of large $N_c$ and large $\lambda$. However, according to the gauge/gravity dictionary, the coupling constants on the gravity side find their counterparts as the central charges on the gauge side. Consequently, the duality between Einstein gravity and conformal field theories (CFTs) is limited to a specific class of CFTs wherein all central charges are equal, thereby lacking the necessary degrees of freedom to accommodate CFTs with distinct central charges \cite{Henningson:1998gx,Henningson:1998ey}. In order to encompass scenarios with finite values of $\lambda$ and $N_c$, which necessitate different central charges, a corresponding dual gravity theory must possess an adequate number of parameters to account for the ratios between central charges. Hence, the extension of the duality conjecture to include conformal field theories with varying central charges requires the consideration of higher-order gravities—those endowed with a richer parameter space capable of accommodating the diverse dual central charges.  Lovelock \cite{Lanczos:1938sf,Lovelock:1971yv,Lov} and various versions of quasi-topological gravity \cite{Oliva:2010eb,Myers:2010ru,Dehghani:2011vu} are among the examples where their additional central charges have been studied \cite{Myers:2010jv,Sinha:2009ev,Buchel:2008vz,Buchel:2008ae,Myers:2008yi}.
	
	The focus of our study revolves around $(n+1)$-dimensional quasi-topological gravity (QTG), which is coupled to an electromagnetic field. This class of theories, along with other higher-order gravitational formulations, introduces additional terms that modify the corresponding action, thus enabling the exploration of valuable dualities \cite{Maldacena:1997re}. An important application arises when the dual gauge theory is constrained by causality requirements. These constraints impose limitations on the coupling constants in the gravitational sector, ensuring the absence of superluminal signals. Notably, these causality constraints align with those imposed by the positivity of energy flux. However, it is worth noting that in the case of gravitational equations of motion beyond second order, this correspondence no longer holds \cite{Hofman:2009ug}. The inclusion of higher-order QTGs, such as cubic, quartic, and quintic formulations, introduces three, four, and five constraints, respectively, which are derived from the requirement of positive energy flux. These constraints play a crucial role in determining the associated coupling constants. To date, there is no evidence suggesting a violation of causality when these higher-order QTGs satisfy these constraints \cite{Oliva:2010eb}.
	
	Similar to Lovelock gravity, QTGs yield field equations that are second-order differential equations with respect to the metric. However, they deviate from Lovelock gravity in dimensions where Lovelock gravity behaves as a topological invariant \cite{Lanczos:1938sf}. It is in these additional dimensions that the requirement for the field equations to be of second order is solely satisfied when the solutions possess spherical symmetry. Nevertheless, when the spherical symmetry is broken, the presence of these additional terms leads to third-order differential equations, hence the term "quasi-topological". Despite this departure from strict topological behavior, the linearized equations governing the propagation of gravitons in an AdS vacuum precisely coincide with those of Einstein gravity. This remarkable correspondence endows QTGs with stable, ghost-free vacua \cite{Myers:2010ru}.
	
In contrast to Lovelock gravity, the QTG encompasses terms of $k$th order curvature, denoted as ${\cal R}^k$, without any dimensional restrictions (except for the case of dimension $2k$) where these terms contribute. This stands in contrast to Lovelock gravity, wherein these terms appear in the field equations only when $k \leq \left[\frac{n}{2}\right]$ \cite{Dehghani:2011vu}.
	
	Cubic, quartic and quintic QTGs (3-, 4- and 5QTG respectively) are studied for example in Refs. \cite{Myers:2010ru,Brenna:2012gp,Ghanaatian:2018nhb}, \cite{Dehghani:2011vu,Ghanaatian:2018gdl,Ghanaatian:2018nhb,Ghanaatian:2014bpa,Chernicoff:2016qrc} and \cite{Bazrafshan:2020fvn} respectively.
	Specially the generic case of 5QTG which construct a quintic quasi-topological gravity in five dimensions is originally studied in \cite{Cisterna:2017umf}.
	
	The corresponding field equations can be solved analytically just up to 4QTG case, whereas for the quintic one, the numerical methods are required to solve the corresponding field equations.
	
		The investigation of 5QTG offers several compelling advantages that serve as the motivation for this study. Firstly, its asymptotically AdS solution provides a broader range of dual conformal field theories (CFTs) compared to lower quartic and cubic formulations. Secondly, the imposition of positive energy flux requirements \cite{Brigante:2008gz,Brigante:2007nu,Ge:2009ac,deBoer:2009pn,Camanho:2009vw} leads to the introduction of five constraints, which in turn determine the values of five coupling constants. Thirdly, while the 4- and 5QTGs possess unique formulations, the uniqueness of the quintic form remains uncertain. For the cubic version, there exist only two possible cubic complete contractions of Weyl tensors, namely $Tr_{(1)}(C^3)$ and $Tr_{(2)}(C^3)$ \cite{Oliva:2010zd}. Similarly, it has been argued in \cite{Dehghani:2011vu} that the 4QTG is unique. However, the uniqueness of the 5QTG is not immediately evident and requires a proper classification of all non-trivial, independent traces of the form $Tr_{(p)}(C^5)$ to be addressed conclusively. Recently, a systematic method for constructing quasi-topological Lagrangians of any order was introduced in \cite{Bueno:2019ycr} through the solution of a recurrence equation.
	
	The existence of QTGs in four dimensions is nonviable. However, this issue can be circumvented by focusing solely on the existence of Static Spherically Symmetric (SSS) solutions, characterized by a single function whose equation involves, at most, second-order derivatives. This leads to the introduction of the broader class of Generalized Quasi-topological Gravities (GQTGs) \cite{Bueno:2017sui,Bueno:2016xff,Hennigar:2017ego,Ahmed:2017jod}, encompassing all QTGs. Notably, intriguing examples of non-trivial GQTGs in four dimensions have been discovered, including the renowned Einsteinian Cubic Gravity \cite{Bueno:2016xff}. Furthermore, proper GQTGs (which do not fall under the QTG classification) have been constructed in dimensions $n \ge 4$, spanning all orders in curvature \cite{Bueno:2019ycr}. GQTGs exhibit linearized equations of motion around maximally symmetric backgrounds \cite{Bueno:2016lrh,Bueno:2017sui,Bueno:2016xff,Hennigar:2017ego,Ahmed:2017jod,Hennigar:2016gkm}, enabling precise calculations of black-hole thermodynamics \cite{Myers:2010ru,Bueno:2017sui,Oliva:2010eb,Cisterna:2017umf,Hennigar:2017ego,Bueno:2017qce}. Moreover, they form a foundation for the entire realm of higher-curvature gravities when considering metric field redefinitions \cite{Bueno:2019ltp}.

	Charged Reissner-Nordström (RN) black holes exhibit a fascinating property in the realm of Einstein gravity: due to their charge, they manifest the behavior of a dual event horizon. An intriguing observation is that, akin to the Lifshitz scenario \cite{Brenna:2011gp}, coupling the RN black holes to the Maxwell field preserves all the advantageous characteristics of QTG. Consequently, considering the advantageous role of charge parameters in the context of the AdS/CFT correspondence, there exists a strong impetus to explore the physical implications stemming from these charged solutions.
	
	The structure of this paper is outlined as follows. In the subsequent section, we introduce the action and derive the corresponding field equations for 5QT gravity, along with the associated black hole solutions. The thermodynamic properties of 5QT black holes are discussed in Section \eqref{sec:Thermo}. Section \eqref{sec:Numeric} is dedicated to the numerical analysis of the field equations. We investigate certain special cases that can be solved analytically in Section \eqref{sec:AnalSol}. Furthermore, the thermal stability of the solutions is examined in Section \eqref{sec:Thermal}. Finally, the concluding section \eqref{sec:Conclusion} presents the main conclusions and additional remarks.
	
	\textcolor{violet}{\section{Field Equations}\label{sec:Action}}
	The action for $(n+1)$-dimensional QTG in the presence of a cosmological constant $\Lambda$ and an electromagnetic field can be expressed as follows:
	\begin{equation}\label{action1}
		I_{G}=\frac{1}{16\pi}\int{d^{n+1}x\sqrt{-g}\big\{-2\Lambda+{\mathcal L}_1+\mu_{2}{\mathcal L}_2+\mu_{3}{\mathcal L}_3+\mu_{4}{\mathcal L}_4+\mu_{5}{\mathcal L}_5-\frac{1}{4}F^2\big\}}.
	\end{equation}
	Here, $F_{\mu\nu} = \partial_{\mu}A_{\nu}-\partial_{\nu}A_{\mu}$ represents the electromagnetic field strength tensor, and $A_{\mu}$ denotes the vector potential. The terms in the action are categorized as follows: the first term corresponds to the cosmological constant, the second term represents the Hilbert-Einstein (HE) terms, and the third term represents the Gauss-Bonnet term ${\mathcal L}_2 = R_{abcd}R^{abcd}-4R_{ab}R^{ab}+R^2$. Furthermore, the fourth term ${{\mathcal L}_3}$ corresponds to the cubic QTG (3QTG), the fifth term ${{\mathcal L}_4}$ corresponds to the quartic QTG (4QTG), and the sixth term ${{\mathcal L}_5}$ corresponds to the quintic QTG (5QTG).
	The explicit forms of the ${{\mathcal L}_3}$ and ${{\mathcal L}_4}$ terms are given by:
	
	\begin{eqnarray}
		{{\mathcal L}_3}&=&
		a_{1}R_{ab}^{cd}R_{cd}^{ef}R_{ef}^{ab}+ a_{2}R_{abcd}R^{abcd}R+a_{3}R_{abcd}R^{abc}{{}_e}R^{de}\nonumber\\
		&&+a_{4}R_{abcd}R^{ac}R^{bd}+a_{5}R_a{{}^b}R_b{{}^c}R_{c}{{}^a}+a_{6}R_a{{}^b}R_b{{}^a}R +a_{7}R^3,
	\end{eqnarray}
	
	\begin{eqnarray}
		{\mathcal{L}_4}&=& b_{1}R_{abcd}R^{cdef}R^{hg}{{}_{ef}}R_{hg}{{}^{ab}}+b_{2}R_{abcd}R^{abcd}R_{ef}{{}^{ef}}+b_{3}RR_{ab}R^{ac}R_c{{}^b}+b_{4}(R_{abcd}R^{abcd})^2\nonumber\\
		&&+b_{5}R_{ab}R^{ac}R_{cd}R^{db}+b_{6}RR_{abcd}R^{ac}R^{db}+b_{7}R_{abcd}R^{ac}R^{be}R^d{{}_e}+b_{8}R_{abcd}R^{acef}R^b{{}_e}R^d{{}_f}\nonumber\\
		&&+b_{9}R_{abcd}R^{ac}R_{ef}R^{bedf}+b_{10}R^4+b_{11}R^2 R_{abcd}R^{abcd}+b_{12}R^2 R_{ab}R^{ab}\nonumber\\
		&&+b_{13}R_{abcd}R^{abef}R_{ef}{{}^c{{}_g}}R^{dg}+b_{14}R_{abcd}R^{aecf}R_{gehf}R^{gbhd},
	\end{eqnarray} 
	and 
	\begin{eqnarray}\label{quintic}
		{\mathcal{L}_5}&=&
		c_{1} R R_{b}^{a} R_{c}^{b} R_{d}^{c} R_{a}^{d}+c_{2} R R_{b}^{a} R_{a}^{b} R_{ef}^{cd} R_{cd}^{ef}+c_{3} R R_{c}^{a} R_{d}^{b} R_{ef}^{cd} R_{ab}^{ef}+c_{4} R_{b}^{a} R_{a}^{b} R_{d}^{c}  R_{e}^{d} R_{c}^{e}\nonumber\\
		&&+c_{5} R_{b}^{a} R_{c}^{b} R_{a}^{c}  R_{fg}^{de} R_{de}^{fg}+c_{6} R_{b}^{a} R_{d}^{b} R_{f}^{c}  R_{ag}^{de} R_{ce}^{fg}+c_{7} R_{b}^{a} R_{d}^{b} R_{f}^{c} R_{cg}^{de} R_{ae}^{fg}+c_{8} R_{b}^{a} R_{c}^{b} R_{ae}^{cd} R_{gh}^{ef} R_{df}^{gh}\nonumber\\
		&&+c_{9} R_{b}^{a} R_{c}^{b} R_{ef}^{cd} R_{gh}^{ef} R_{ad}^{gh}+c_{10} R_{b}^{a} R_{c}^{b} R_{eg}^{cd} R_{ah}^{ef} R_{df}^{gh}+c_{11} R_{c}^{a} R_{d}^{b} R_{ab}^{cd} R_{gh}^{ef} R_{ef}^{gh}+c_{12} R_{c}^{a} R_{d}^{b} R_{ae}^{cd} R_{gh}^{ef} R_{bf}^{gh}\nonumber\\
		&&+c_{13} R_{c}^{a} R_{d}^{b} R_{ef}^{cd} R_{gh}^{ef} R_{ab}^{gh}+c_{14} R_{c}^{a} R_{d}^{b} R_{eg}^{cd} R_{ah}^{ef} R_{bf}^{gh}+c_{15} R_{c}^{a} R_{e}^{b} R_{af}^{cd} R_{gh}^{ef} R_{bd}^{gh}+c_{16} R_{b}^{a} R_{ad}^{bc} R_{fh}^{de} R_{ci}^{fg} R_{eg}^{hi}\nonumber\\
		&&+c_{17} R_{b}^{a} R_{de}^{bc} R_{cf}^{de} R_{hi}^{fg} R_{ag}^{hi}+c_{18} R_{b}^{a} R_{df}^{bc} R_{ac}^{de} R_{hi}^{fg} R_{eg}^{hi}+c_{19} R_{b}^{a} R_{df}^{bc} R_{ah}^{de} R_{ei}^{fg} R_{cg}^{hi}+c_{20} R_{b}^{a} R_{df}^{bc} R_{gh}^{de} R_{ei}^{fg} R_{ac}^{hi}\nonumber\\
		&&+c_{21} R_{cd}^{ab} R_{eg}^{cd} R_{ai}^{ef} R_{fj}^{gh}R_{bh}^{ij}+c_{22} R_{ce}^{ab} R_{af}^{cd} R_{gi}^{ef} R_{bj}^{gh}R_{dh}^{ij}+c_{23} R_{ce}^{ab} R_{ag}^{cd} R_{bi}^{ef} R_{fj}^{gh}R_{dh}^{ij}\nonumber\\
		&&+c_{24} R_{ce}^{ab} R_{fg}^{cd} R_{hi}^{ef} R_{aj}^{gh}R_{bd}^{ij}. 
	\end{eqnarray}
	The coefficients $a_{i}$, $b_{i}$, and $c_{i}$ associated with these terms can be found in Tables \eqref{tab:cubic}, \eqref{tab:quart}, and \eqref{tab:quint} in the Appendix.
	
	We consider the metric in the following form:
	\begin{eqnarray}\label{metr}
		ds^2=-\frac{r^2}{L^2}f(r)dt^2+\frac{L^2}{r^2 g(r)}dr^2+\frac{r^2}{L^2} d\Omega_k^2.
	\end{eqnarray}
	Here, $L$ is a scale factor that is associated with the cosmological constant $\Lambda$. The metric functions $f(r)$ and $g(r)$ define the properties of the metric, and $d\Omega_k^2$ represents a constant curvature hypersurface with a Euclidean metric. The specific form of $d\Omega_k^2$ is given by:
	\begin{equation}
		d\Omega_k^2=\left\{
		\begin{array}{ll}
			$$d\theta^{2}_{1}+\sum_{i=2}^{n-1}\prod_{j=1}^{i-1} \sin^2 \theta_{j} d\theta_{i}^2$$,\quad \quad\quad\quad \quad\quad\quad\quad\quad\quad  \ {k=1,}\quad &  \\ \\
			$$\sum_{i=1}^{n-1} d\phi_{i}^{2}$$,\quad\quad\quad\quad\quad\quad \quad\quad\quad\quad\quad\quad\quad\quad\quad\quad\quad\quad  \ {k=0,}\quad &  \\ \\
			$$d\theta^{2}_{1}+\sinh^2 \theta_1 d\theta_{2}^{2}+\sinh^2 \theta_{1} \sum_{i=3}^{n-1}\prod_{j=2}^{i-1} \sin^2 \theta_{j} d\theta_{i}^2$$, \quad{k=-1~.}\quad &
		\end{array}
		\right.
	\end{equation}
	Here, the parameter $k$ takes values $1$, $0$, or $-1$, which correspond to spherical, flat, and hyperbolic geometries, respectively.
	
	For the vector potential, we consider the following form:
	\begin{eqnarray}\label{h1}
		A_{\mu}=h(r)\delta_{\mu}^{0}~.
	\end{eqnarray}
	This choice of vector potential leads to static solutions.
	
	By considering the metric \eqref{metr} and performing integration by parts on the action \eqref{action1}, we obtain the following action:
	\begin{eqnarray}\label{Act2} 
		I_G &=& \frac{(n-1)}{16\pi L^2}\int d^{n} x\int dr \sqrt{\frac{f}{g}}\bigg\{\bigg[r^n\bigg(\hat{\mu}_{0}-\Psi+\hat{\mu}_{2}\Psi^2+\hat{\mu}_{3}\Psi^3+\hat{\mu}_{4}\Psi^4+\hat{\mu}_{5}\Psi^5\bigg)\bigg]^{'} \nonumber\\
		&&+\frac{r^{n-1}g}{2(n-1)f}(h^{'})^2\bigg\}~.
	\end{eqnarray}
	In this expression, $\Psi = \left(g-\frac{L^2}{r^2}k\right)$ and $(')$ denotes the derivative with respect to $r$. Additionally, we employ dimensionless coefficients $\hat{\mu}_i$, which are defined as:
	\begin{eqnarray}
		\Lambda=-\frac{n(n-1)\hat{\mu}_{0}}{2L^2}~,
	\end{eqnarray}
	\begin{eqnarray}
		\mu_{2}=\frac{\hat{\mu}_{2} L^2}{(n-2)(n-3)},
	\end{eqnarray}
	\begin{eqnarray}
		\mu_{3}=\frac{8(2n-1)\hat{\mu}_{3}L^4}{(n-2)(n-5)(3n^2-9n+4)},
	\end{eqnarray}
	\begin{eqnarray}
		\mu_{4}=\frac{\hat{\mu}_{4}L^6}{n(n-1)(n-3)(n-7)(n-2)^2(n^5-15n^4+72n^3-156n^2+150n-42)},
	\end{eqnarray}
	\begin{eqnarray}\label{mu5}
		\mu_{5}&=&\frac{\hat{\mu}_{{5}}{L}^{8}}{(n-3)(n-9)(n-2)^2 } \Big(8\,{n}^{12}+26\,{n}^{11}-1489\,{n}^{10}+11130\,{n}^{9}-26362\,{n}^{8}\nonumber\\
		&&-
		75132\,{n}^{7}+705657\,{n}^{6}-2318456\,{n}^{5}+4461054\,{n}^{4}-
		5484168\,{n}^{3}\nonumber\\
		&&+4290516\,{n}^{2}-1968224\,n+405376\Big)^{-1}.
	\end{eqnarray}
	The field equations in $(n+1)$ dimensions can be obtained by varying the action with respect to $f(r)$, $g(r)$, and $h(r)$, and they take the following form:
	\begin{align}
		&- \frac{ r^{10}}{4 f}\left[ g \left(h^{'} \right)^2  \right] = - \frac{L^2}{2}r^{10}n(n-1)\Big[-\frac{2\Lambda}{n(n-1)}+g+\hat{\mu}_{2}g^2+\hat{\mu}_{3}g^3+\hat{\mu}_{4}g^4+\hat{\mu}_{5}g^5\Big]\nonumber\\& - \frac{\hat{\mu}_{2}L^2}{4}\Big[ 2 r^6 L^4 k^2(n^2-5n+4)-4g r^8 L^2 k(n^2-3n+2)\Big]\nonumber\\
		&-\frac{\hat{\mu}_{3}L^2}{4}\Big[-2 r^4 L^6 k(n^2-7n+6)+6r^6 L^4 k^2(n^2-5n+4)g-6g^2 r^8 l^2 k(n^2-3n+2)\Big]\nonumber\\
		&-\frac{\hat{\mu}_{4}L^2}{4}\Big[2r^2 L^8 k^2(n^2-9n+8)-8gr^4 L^6k(n^2-7n+6)+12g^2r^6L^4k^2(n^2-5n+4) \nonumber \\
		&-8r^8L^2kg^3(n^2-3n+2)\Big]-\frac{\hat{\mu}_{5}L^2}{4}\Big[-2r^{10}k(n^2-11n+10)+10gr^2L^8k^2(n^2-10n+8)\nonumber\\ &-20g^2r^4L^6k(n^2-7n+6)+20g^3r^6L^4k^2(n^2-5n+4)-10L^2kr^8g^4(n^2-3n+2)\Big]\nonumber\\
		&-(\ln f)^{\prime}\frac{L^2(n-1)}{2}\Big[r^{11}(g+2\hat{\mu}_{2}g^2+3\hat{\mu}_{3}g^3+4\hat{\mu}_{4}g^4+5\hat{\mu}_{5}g^5)-2\hat{\mu}_{2}r^9L^2kg-6\hat{\mu}_{3}kL^2g^2r^9\nonumber\\& +3r^7L^4k^2\hat{\mu}_{3}g-12\hat{\mu}_{4}r^9g^3kL^2+12\hat{\mu}_{4}r^7g^2k^2L^4-4L^6kr^5g\hat{\mu}_{4}-20r^9L^2kg^4\hat{\mu}_{5} \nonumber\\& +30k^2L^4r^7\hat{\mu}_{5}g^3-20r^5L^6kg^2\hat{\mu}_{5}+5r^3L^8k^2g\hat{\mu}_{5}\Big]~, \label{fieldequations_initial}\\
		& \frac{ r^{(n-1)}}{2 f}\left[ g \left( h^{'}\right)^2  \right] = \Big( (n-1) r^n \left[ \hat{\mu}_{0}%
		- \Psi + \hat{\mu}_{2} \Psi^2 + \hat{\mu}_{3} \Psi^3+ \hat{\mu}_{4} \Psi^4+ \hat{\mu}_{5} \Psi^5 \right] \Big)^{\prime}~,
		\label{fieldequations_initial2}\\
		&2rh^{\prime \prime }- r h^{\prime }\left[ (\ln f)^{\prime }-(\ln%
		g)^{\prime }\right] +2(n-1)h^{\prime }=0~.
		\label{fieldequations_final}
	\end{align}
	By using equations \eqref{metr} and \eqref{h1}, and assuming $f(r) = N^2(r) g(r)$, the above equations can be simplified as:
	\begin{eqnarray}
		(1-2\hat{\mu}_{2} \Psi+3\hat{\mu}_{3}\Psi^2+4\hat{\mu}_{4}\Psi^3+5\hat{\mu}_{5}\Psi^4)N^{'}=0~, \label{finalN}\\
		\left( (n-1) r^n \left[ \hat{\mu}_{0}%
		- \Psi + \hat{\mu}_{2} \Psi^2 + \hat{\mu}_{3} \Psi^3+ \hat{\mu}_{4} \Psi^4+ \hat{\mu}_{5} \Psi^5 \right] \right)^{\prime}%
		&=& \frac{ r^{(n-1)}}{2}\left[  \left(\frac{\left( h \right)^\prime}{N}\right)^2  \right],
		\label{finalg}
	\end{eqnarray}
	\begin{eqnarray}
		\left( {r^{(n-1)}} \left( h \right)^\prime  \right)^\prime  &=&0~,
		\label{finalh}
	\end{eqnarray}
	where
	\begin{eqnarray}
		h(r)&=&-\sqrt{\frac{2(n-1)}{n-2}}\frac{q}{r^{n-2}}~.  \label{hr}
	\end{eqnarray}
	Here, the parameter $q$ is associated with the black hole charge, which can be determined using Gauss' law as:
	\begin{eqnarray}
		Q &=&\frac{%
			\sqrt{2(n-1)(n-2)}}{16\pi } q~.
	\end{eqnarray}
	To find the solutions, we start with equation \eqref{finalN}. This equation indicates that $N(r)$ needs to be constant, and thus we can choose $N(r) = 1$. By substituting this choice and combining equations \eqref{hr} and \eqref{finalg}, we obtain:
	\begin{eqnarray}\label{equasli}
		\hat{\mu}_{5} \Psi^5+\hat{\mu}_{4} \Psi^4+\hat{\mu}_{3} \Psi^3+\hat{\mu}_{2} \Psi^2-\Psi+\kappa=0,
	\end{eqnarray}
	where
	\begin{eqnarray}\label{kappa}
		\kappa=\hat{\mu}_{0}-\frac{m}{r^{n}}+\frac{q^{2}}{r^{2n-2}}.
	\end{eqnarray}
	Here, the constant of integration $m$ is related to the black hole mass. The geometrical mass of the black hole can be calculated as:
	\begin{eqnarray}\label{geomass}
		m=r_{+}^{n}\bigg(\hat{\mu}_{0}-k\frac{L^2}{r_{+}^2}+\hat{\mu}_{2} k^2\frac{L^4}{r_{+}^4}-\hat{\mu}_{3} k^3\frac{L^6}{r_{+}^6}+\hat{\mu}_{4} k^4\frac{L^8}{r_{+}^8}-\hat{\mu}_{5} k^5\frac{L^{10}}{r_{+}^{10}}+\frac{q^2}{r_{+}^{(2n-2)}}\bigg),
	\end{eqnarray}
	where $r_+$ represents the positive root of the equation $f(r_+) = 0$, which corresponds to the outermost radius of the black hole.

	\textcolor{violet}{\section{Thermodynamics }\label{sec:Thermo}}
	The study of the thermodynamic properties of black holes has significantly contributed to our understanding of quantum gravity, particularly through the development of the holographic principle. In this section, we focus on investigating the stability of black holes by examining their statistical mechanics.
	
	To calculate the entropy of black holes, we employ the Iyer-Wald formula \cite{Iyer:1994ys}, given by:
	\begin{eqnarray}\label{entropy}
		s=-2\pi \oint d^{n-1} x \sqrt{\tilde{g}} \frac{\partial \mathcal L}{\partial R_{abcd}}\hat{\epsilon}{ab}\hat{\epsilon}{cd}.
	\end{eqnarray}
	Here, $\hat{\epsilon}_{ab}$ represents the binormal of the black hole horizon, and $\tilde{g}$ denotes the determinant of the induced metric on the horizon.
	
	The Lagrangian $\mathcal{L}$ encompasses the Hilbert-Einstein, Gauss-Bonnet, and quasi-topological terms of order 3, 4, and 5. By defining $Y=\frac{\partial \mathcal{L}}{\partial R_{abcd}}\hat{\epsilon}_{ab}\hat{\epsilon}_{cd}$, the values of $Y$ in the 4- and 5QTG can be found in references \cite{Dehghani:2011vu,Myers:2010ru}, respectively. To calculate the corresponding terms for 5QTG, one needs to vary all 24 terms present in the Lagrangian \eqref{quintic}. Here, we provide a few examples of these terms \cite{Bazrafshan:2019oan}:
	\begin{eqnarray}
		Y_{4}&=& 2R_{d}^{c}R_{e}^{d}R_{c}^{e}(R^{t}_{t}+R^{r}_{r})+3(R_{e}^{t}R_{t}^{e}+R_{e}^{r}R_{r}^{e})R_{b}^{a}R_{a}^{b},\nonumber\\
		Y_{8}&=& R_{f}^{c}R_{cg}^{de}(R_{d}^{r}R_{re}^{fg}+R_{d}^{t}R_{te}^{fg})+R_{f}^{c}R_{ae}^{fg}(R_{r}^{a}R_{cg}^{re}+R_{t}^{a}R_{cg}^{te})+R_{b}^{a}R_{d}^{b}(R_{rg}^{de}R_{ae}^{rg}+R_{tg}^{de}R_{ae}^{tg})\nonumber\\
		&&+R_{b}^{a}(R_{r}^{b}R_{f}^{r}R_{at}^{ft}+R_{t}^{b}R_{f}^{t}R_{ar}^{fr}-R_{r}^{b}R_{f}^{t}R_{at}^{fr}-R_{t}^{b}R_{f}^{r}R_{ar}^{ft})+R_{d}^{b}(R_{b}^{r}R_{r}^{c}R_{ct}^{dt}+R_{b}^{t}R_{t}^{c}R_{cr}^{dr}\nonumber\\
		&&-R_{b}^{t}R_{r}^{c}R_{ct}^{dr}-R_{b}^{r}R_{t}^{c}R_{cr}^{dt}),\nonumber\\
		Y_{23}&=& R_{fj}^{gh}R_{dh}^{ij}(R_{rg}^{rd}R_{ti}^{tf}
		+R_{tg}^{td}R_{ri}^{rf}-R_{rg}^{td}R_{ti}^{rf}
		-R_{tg}^{rd}R_{ri}^{tf})+R_{bi}^{ef} (R_{re}^{rb}R_{fj}^{th}R_{th}^{ij}+R_{te}^{tb}R_{fj}^{rh}R_{rh}^{ij}\nonumber\\
		&&-R_{re}^{tb}R_{fj}^{rh}R_{th}^{ij}-
		R_{te}^{rb}R_{fj}^{th}R_{rh}^{ij})+R_{ag}^{cd}(R_{cr}^{ar}R_{tj}^{gh}R_{dh}^{tj}+R_{ct}^{at}R_{rj}^{gh}R_{dh}^{rj}-R_{cr}^{at}R_{tj}^{gh} R_{dh}^{rj}\nonumber\\
		&&-R_{ct}^{ar}R_{rj}^{gh}R_{dh}^{tj})+
		R_{ce}^{ab}(R_{ar}^{cd}R_{bi}^{er}R_{dt}^{it}
		+R_{at}^{cd}R_{bi}^{et}R_{dr}^{ir}-R_{ar}^{cd}
		R_{bi}^{et}R_{dt}^{ir}-R_{at}^{cd}R_{bi}^{er}R_{dr}^{it})\nonumber\\
		&&+R_{ce}^{ab}(R_{ag}^{cr}R_{br}^{ef}
		R_{ft}^{gt}+R_{ag}^{ct}R_{bt}^{ef}R_{fr}^{gr}
		-R_{ag}^{ct}R_{br}^{ef}R_{ft}^{gr}-R_{ag}^{cr}
		R_{bt}^{ef}R_{fr}^{gt}), \nonumber\\
		Y_{24}&=& (R_{fg}^{rd}R_{hi}^{tf}R_{rj}^{gh}R_{td}^{ij}
		+R_{fg}^{td}R_{hi}^{rf}R_{tj}^{gh}R_{rd}^{ij}
		-R_{fg}^{td}R_{hi}^{rf}R_{rj}^{gh}R_{td}^{ij}
		-R_{fg}^{rd}R_{hi}^{tf}R_{tj}^{gh}R_{rd}^{ij})\nonumber\\
		&&+
		(R_{re}^{ab}R_{hi}^{er}R_{aj}^{th}R_{bt}^{ij}+
		R_{te}^{ab}R_{hi}^{et}R_{aj}^{rh}R_{br}^{ij}-R_{re}^{ab}R_{hi}^{et}R_{aj}^{rh}R_{bt}^{ij}
		-R_{te}^{ab}R_{hi}^{er}R_{aj}^{th}R_{br}^{ij})\nonumber\\
		&&+(R_{cr}^{ab}R_{tg}^{cd}R_{aj}^{gr}R_{bd}^{tj}+
		R_{ct}^{ab}R_{rg}^{cd}R_{aj}^{gt}R_{bd}^{rj}
		-R_{cr}^{ab}R_{tg}^{cd}R_{aj}^{gt}R_{bd}^{rj}
		-R_{ct}^{ab}R_{rg}^{cd}R_{aj}^{gr}R_{bd}^{tj})\nonumber\\
		&&+(R_{ce}^{rb}R_{fr}^{cd}R_{ti}^{ef}R_{bd}^{it}+
		R_{ce}^{tb}R_{ft}^{cd}R_{ri}^{ef}R_{bd}^{ir}
		-R_{ce}^{tb}R_{fr}^{cd}R_{ti}^{ef}R_{bd}^{ir}
		-R_{ce}^{rb}R_{ft}^{cd}R_{ri}^{ef}R_{bd}^{it})\nonumber\\
		&&+(R_{ce}^{ar}R_{fg}^{ct}R_{hr}^{ef}R_{at}^{gh}+
		R_{ce}^{at}R_{fg}^{cr}R_{ht}^{ef}R_{ar}^{gh}-R_{ce}^{at}R_{fg}^{cr}R_{hr}^{ef}R_{at}^{gh}
		-R_{ce}^{ar}R_{fg}^{ct}R_{ht}^{ef}R_{ar}^{gh}).
	\end{eqnarray}
	These expressions illustrate the calculation of a few terms within the Lagrangian for 5QTG, and additional terms can be determined by considering the remaining terms in the Lagrangian \eqref{quintic}.
	
	Finally, considering the entropy density as $S=s/V_{n-1}$, for the entropy in the quintic case, we obtain:
	\begin{eqnarray}
		S_{5}= \frac{5(n-1)L^8k^4\hat{\mu}_{5}}{4(n-9)}r_{+}^{n-9}.
	\end{eqnarray}
	Here, $s$ represents the entropy and $V_{n-1}$ denotes the volume of the $(n-1)$-dimensional space. 
Taking into account all the contributions, the total value of the entropy $S$ can be determined as:
\begin{eqnarray}\label{enf}
	S=\frac{r_{+}^{n-1}}{4}\bigg(1+2k\hat{\mu}_{2} \frac{(n-1)L^2}{(n-3)r_{+}^2}-3k^2\hat{\mu}_{3} \frac{(n-1)L^4}{(n-5)r_{+}^4}+4k^3\hat{\mu}_{4} \frac{(n-1)L^6}{(n-7)r_{+}^6}-5k^4\hat{\mu}_{5} \frac{(n-1)L^8}{(n-9)r_{+}^8}\bigg).
\end{eqnarray}
Here, $r_+$ represents the radial coordinate of the outermost horizon of the black hole, which is the positive root of the equation $f(r_+) = 0$. The above equation provides the entropy of the black hole, where setting $\hat{\mu}_2=\hat{\mu}_3=\hat{\mu}_4=\hat{\mu}_5=0$ recovers the entropy for Hilbert-Einstein gravity. Consequently, higher-order theories of the curvature tensor introduce correction terms to the Hilbert-Einstein entropy.

The temperature of the event horizon can be calculated using the standard method of analytic continuation of the metric:
\begin{eqnarray}\label{T+}
	T_+ &=& \Big(\frac{r^2 g'}{4 \pi L^2}\Big)_{r_+}\nonumber\\
	&=&\frac {1}{4\pi r_+( {r}_+^{8}+2\,\hat{\mu}_{2}kL^2{r}_+^{6}+3\,\hat{\mu}_{3}{k}^{2}L^4{r}_+^{4}+4\,\hat{\mu}_{4}{k}^{3}{r}_+^{2}L^6+5\,\hat{\mu}_{5}{k}^{4}L^8)}\nonumber\\
	&&\times \Big[ -\left( n-2 \right){q}^{2}r_+^{-2(n-6)}-\left( n-10 \right){k}^{5}\hat{\mu}_{5}L^{10}+\left( n-8
	\right){k}^{4} \hat{\mu}_{4} L^{8}{r}_+^{2} \nonumber\\&& -\left( n-6 \right){k}^{3}\hat{\mu}_{3} L^{6}{r}_+^{4}+\left(n-
	4\right){k}^{2}\hat{\mu}_{2} L^{4}{r}_+^{6}+\left( n-2 \right) k L^{2}  {r}_+^{8}+n \hat{\mu}_{0}{r
	}_+^{10} \Big].
\end{eqnarray}
The electric potential $\Phi$ at infinity with respect to the event horizon is defined as:
\begin{eqnarray}
	\Phi=A_{\mu }\chi ^{\mu }\left| _{r\rightarrow \infty }-A_{\mu }\chi
	^{\mu }\right| _{r=r_{+}},  \label{Pot}
\end{eqnarray}
where $\chi=\partial_{t}$ represents the null generator of the horizon. By applying equation \eqref{Pot}, the electric potential can be obtained as:
\begin{eqnarray}\label{potential}
	\Phi=\sqrt{\frac{2(n-1)}{n-2}}\frac{q}{ r_{+}^{n-2}}.
\end{eqnarray}
To calculate the mass of the black hole, the subtraction method can be employed. Writing the metric in the following form:
\begin{eqnarray}\label{metr2}
	ds^2 = -W^2(r)dt^2 + \frac{dr^2}{V^2(r)} + r^2 d\Omega^2~,
\end{eqnarray}
the quasi-local mass per unit volume $V_n$ can be calculated as:
\begin{eqnarray}
	{\cal M} = r W(r)(V_0(r)-V(r))~.
\end{eqnarray}
Here, $V_0(r)$ represents the zero of energy. In the limit $r \rightarrow \infty$, ${\cal M}$ yields the Arnowitt-Deser-Misner mass. Thus, the mass per unit volume can be obtained by expressing the metric \eqref{metr} in the form of equation \eqref{metr2} as:
\begin{eqnarray}\label{massM}
	M = \dfrac{(n-1)m}{16 \pi L^2}.
\end{eqnarray}

We define the corresponding intensive parameters $T$ and $\Phi$, associated respectively with the entropy $s$ and charge $Q$, for the mass $M(s,Q)$ as:
\begin{eqnarray}
	T=\Big(\frac{\partial M}{\partial s}\Big)_Q, ~~~ \Phi = \Big(\frac{\partial M}{\partial Q}\Big)_s.
\end{eqnarray}
These parameters are given by equations \eqref{T+} and \eqref{potential}, which satisfy the first law of black hole thermodynamics:
\begin{eqnarray}
	dM = T ds + \Phi dQ.
\end{eqnarray}
In the extended phase space, where the cosmological constant $\Lambda$, Gauss-Bonnet, cubic, quartic, and quintic coefficients ($\hat{\mu}_2$ to $\hat{\mu}_5$) are considered as thermodynamic variables, the first law of thermodynamics is generalized. The cosmological constant plays the role of thermodynamic pressure:
\begin{eqnarray}\label{PVBH}
	P = -\frac{\Lambda}{8\pi}~,~~~~
	V = \frac{r_+^n}{n}~.
\end{eqnarray}
Here, $V$ represents the volume conjugate to the pressure $P$ \cite{Dolan:2011xt}, and the specific volume is defined as $v=4r_+/(n-1)$. Therefore, the first law of thermodynamics in the extended phase space can be expressed as:
\begin{eqnarray}\label{1stlawext}
	dM = Tds + \Phi dQ + V dP + \Psi_2 d\hat{\mu}_2 + \Psi_3 d\hat{\mu}_3 + \Psi_4 d\hat{\mu}_4 + \Psi_5 d\hat{\mu}_5~,
\end{eqnarray}
where $\Psi_i$ are conjugate variables to $\hat{\mu}_i$. By using equations \eqref{geomass} and \eqref{massM}, one can calculate $\Psi_i$ as follows:
\begin{eqnarray}\label{Psiofmu}
	\Psi_2 &=& \frac{\partial M}{\partial \hat{\mu}_2} = \frac{k^2 (n-1) r_+^{n-4}}{16 \pi }-\frac{k (n-1) r_+^{n-3}}{2 (n-3)}T_+~,\\
	\Psi_3 &=& \frac{\partial M}{\partial \hat{\mu}_3} = \frac{3 k^2 (n-1)  r_+^{n-5}}{4 (n-5)}-\frac{k^3 (n-1) r_+^{n-6}}{16 \pi }T_+~,\\
	\Psi_4 &=& \frac{\partial M}{\partial \hat{\mu}_4} = \frac{k^4 (n-1) r_+^{n-8}}{16 \pi }-\frac{k^3 (n-1) r_+^{n-7}}{n-7}T_+~,\\
	\Psi_5 &=& \frac{\partial M}{\partial \hat{\mu}_5} = \frac{5 k^4 (n-1) r_+^{n-9}}{4 (n-9)}-\frac{k^5 (n-1) r_+^{n-10}}{16 \pi }T_+~.
\end{eqnarray}
	
	\textcolor{teal}{\subsection*{\textit{Critical Behavior}}}
	To examine the critical behavior of QTG, it is necessary to obtain the $P-v$ isotherm for the corresponding black hole. By utilizing equations \eqref{T+} and \eqref{PVBH}, we can derive the equation of state as follows:
	\begin{eqnarray}\label{eqos}
		P &=& \frac{T_+}{v}-\frac{(n-2)k}{\pi(n-1)v^2} +\frac{2^{4n-1}\pi Q^2}{\big[(n-1)v\big]^{2n-2}}\nonumber\\ &&+
		\sum_{j=2}^{5}\Bigg[\frac{(-1)^j j 16^{j-1}~k^{j-1}}{(n-1)^{2j-2}~ v^{2j-1}}  ~\Bigg( T_+ - \frac{(n-2j)k}{j \pi (n-1) v} \Bigg)~\hat{\mu}_j\Bigg]~.
	\end{eqnarray}
	
	This equation represents the equation of state, where $P$ denotes the pressure, $T_+$ is the temperature, $v$ represents the specific volume, $k$ is a constant, $Q$ is the charge parameter, and $\hat{\mu}_j$ represents additional parameters associated with the quasi-topological terms. The equation \eqref{eqos} captures the relationship between pressure, volume, temperature, and the various parameters characterizing the black hole, enabling the study of its critical behavior.
	
\begin{figure}
		\includegraphics[width=7.5cm, height=6cm]{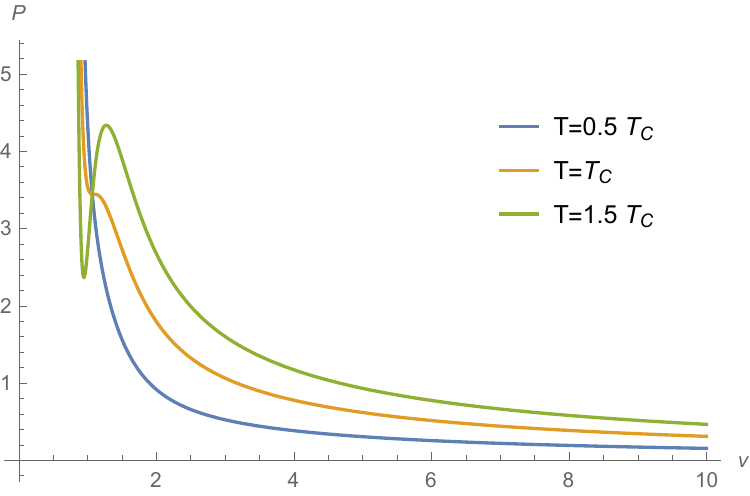}
		\vspace{-0.5cm} 
		\includegraphics[width=7.5cm, height=6cm]{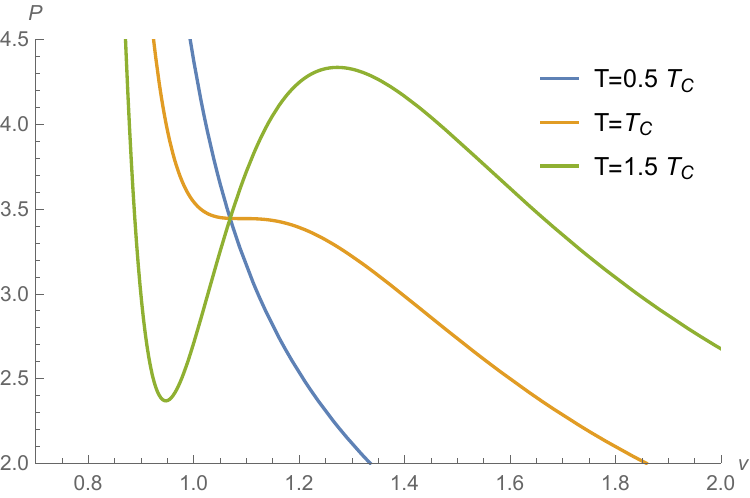}
		\caption{Plot of Pressure $P$ versus specific volume $v$ for different values of $T$ and $k=1$, $n=4$, $Q=0.2$, $\hat{\mu}_2=-0.01$, $\hat{\mu}_3=-0.5$, $\hat{\mu}_4=-0.5$ and $\hat{\mu}_5=-0.1$. With these parameters the critical pressure, specific volume and temperature are obtained as $P_C = 3.44$, $v_C = 1.09$ and $T_C = 3.11$. }
		\label{fig:crit}
	\end{figure}

	To determine the critical point, the following system of equations needs to be solved:
	\begin{eqnarray}\label{crit}
		\frac{\partial P}{\partial v} = 0~, \quad \frac{\partial^2 P}{\partial v^2} = 0~.
	\end{eqnarray}
	The critical values of pressure, specific volume, and temperature, denoted as ($P_C$, $v_C$, $T_C$), satisfy these conditions. For the hyperbolic case with $k=-1$, it is found that the equations \eqref{crit} do not have a solution, indicating the absence of a critical point in this scenario. However, in the spherical case with $k=1$, a critical point can be identified for the given parameters, as depicted in Fig. \eqref{fig:crit}.
	
	The pressure ($P$) as a function of specific volume ($v$) displays notable distinctions in its behavior across different temperature regimes, namely $T<T_C$, $T=T_C$, and $T>T_C$, despite their asymptotic resemblances. Figure \eqref{fig:crit} illustrates the behavior near the critical point, focusing on the right panel. Remarkably, at the critical point where the temperature assumes its critical value $T=T_C$ (depicted by the orange curve), the conditions outlined by \eqref{crit} are fulfilled.
		
\begin{figure}[h]
		\centerline{\includegraphics[width=0.7\textwidth]{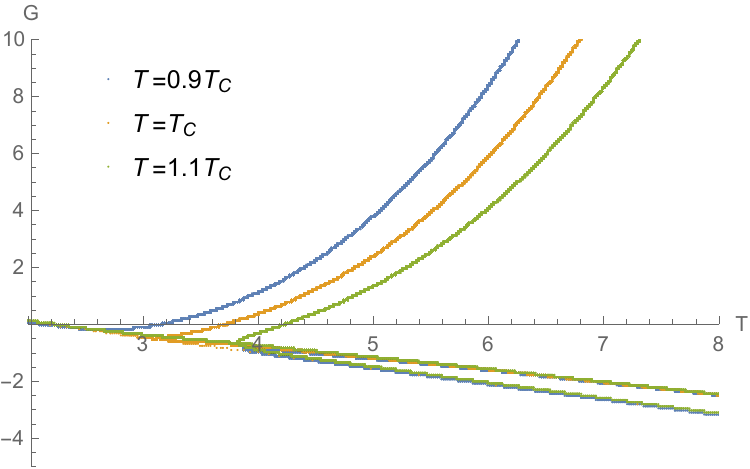}}
		\caption{Plot of $G$ vs $T$ near the critical point $(T_C, v_C, P_C) = (3.11, 1.09, 3.44)$.}
		\label{fig:GT}
	\end{figure}

	To investigate the phase transition, the Gibbs function ($G$) is calculated using the expression:
	\begin{eqnarray}\label{Gibbs}
		G = M - T S~.
	\end{eqnarray}
	The plot of $G$ as a function of temperature ($T$) is shown in Figure \ref{fig:GT}.
	
	It is evident that in the regime where $T<T_C$, the absence of a swallowtail indicates the absence of a first-order phase transition. Instead, this suggests the occurrence of a zeroth-order phase transition, wherein a transition from a small black hole to a large one takes place.
	
	\textcolor{violet}{\section{Numerical Analyses}\label{sec:Numeric}}
Given the absence of an analytical solution for the quintic field equation \eqref{equasli}, numerical methods are employed to generate a plot illustrating the behavior of the function $f(r)$. The various scenarios, characterized by $\Lambda<0$, $\Lambda>0$, and $\Lambda=0$, correspond to asymptotically Anti-de Sitter (AdS), de Sitter (dS), and flat spacetime solutions, respectively.
	
	\textcolor{teal}{\subsection*{\textit{Asymptotically AdS spacetimes}}}
To thoroughly investigate spacetimes that are asymptotically Anti-de Sitter (AdS), it is crucial to consider the condition $\lim_{r\rightarrow \infty} f(r)=1$ to ensure the desired behavior. By incorporating this limit, we can determine the value of the cosmological constant through the equation
\begin{eqnarray}\label{lambda_ads}
	\Lambda =\frac{n(n-1)}{2L^2}(\hat{\mu}_5 + \hat{\mu}_4 + \hat{\mu}_3 + \hat{\mu}_2 - 1),
\end{eqnarray}
where $\hat{\mu}_5$, $\hat{\mu}_4$, $\hat{\mu}_3$, and $\hat{\mu}_2$ represent coefficients. In order to maintain the characteristic of asymptotically AdS spacetimes, the cosmological constant $\Lambda$ must be negative, imposing the condition $\hat{\mu}_5 + \hat{\mu}_4 + \hat{\mu}_3 + \hat{\mu}_2 < 1$. Complying with this condition, the Reissner-Nordström solutions for the radius of a given black hole can be derived.	
	


\begin{figure}[h]
		\centerline{\includegraphics[width=0.7\textwidth]{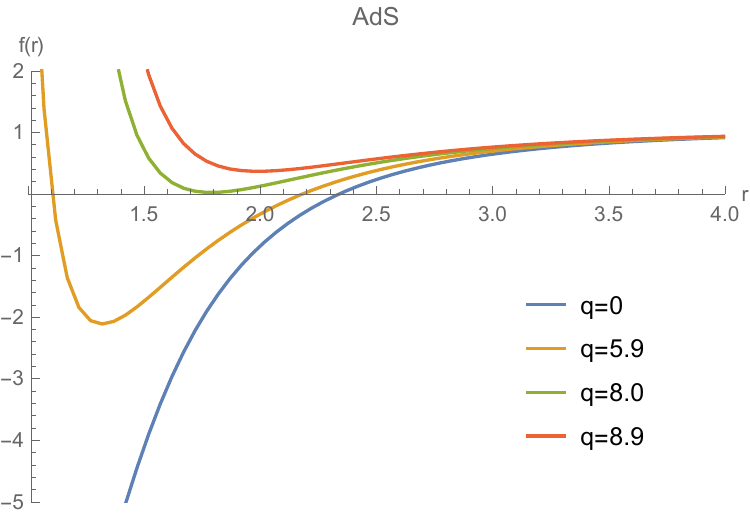}}
		\caption{$f(r)$ for asymptotically AdS spacetime versus $r$ for different values of $q$ with $m=30$, $k=0$, $\hat{\mu}_2=0.04$, $\hat{\mu}_3=-0.001$, $\hat{\mu}_4=-0.0002$, $\hat{\mu}_5=-0.0001$.}
		\label{fig:AdsfvsrforQ}
	\end{figure}

\begin{figure}[h]
		\centerline{\includegraphics[width=0.7\textwidth]{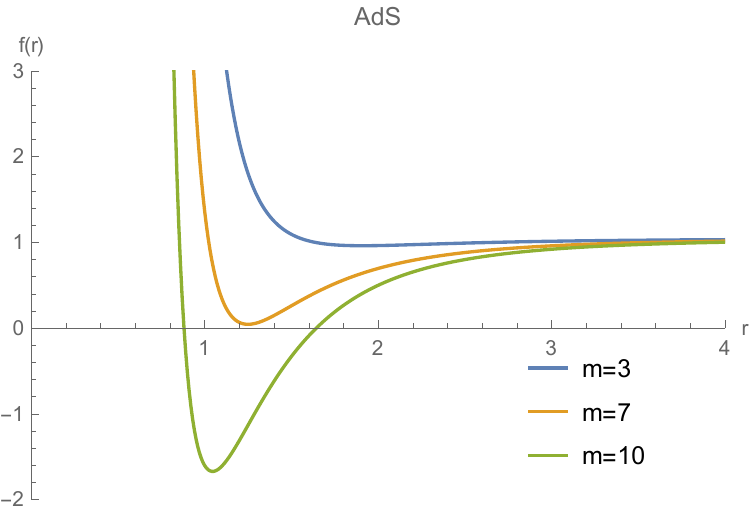}}
		\caption{$f(r)$ for asymptotically AdS spacetime versus $r$ for different values of $m$ with $q=2.7$, $k=0$, $\hat{\mu}_2=0.04$, $\hat{\mu}_3=-0.001$, $\hat{\mu}_4=-0.0002$, $\hat{\mu}_5=-0.0001$.}
		\label{fig:AdsfvsrforM}
	\end{figure}

	
	Figure \ref{fig:AdsfvsrforQ} showcases the plot of $f(r)$ as a function of $r$ for various values of $q$ within the context of 5QTG. Notably, it is apparent that two critical points, namely $q_{min}$ and $q_{ext}$, can be identified. Under the condition of fixed parameters, including $m$, $k$, $\hat{\mu}_2$, $\hat{\mu}_3$, $\hat{\mu}_4$, and $\hat{\mu}_5$, the following classifications can be made: for $q < q_{min}$, there exists a non-extremal black hole; for $q_{min} < q < q_{ext}$, a black hole with two horizons is present; for $q = q_{ext}$, an extremal black hole is observed; and for $q > q_{ext}$, a naked singularity emerges. It is noteworthy that $q_{ext}$ takes on a value of $8.0$.
	Figure \eqref{fig:AdsfvsrforM} illustrates the dependence of the black hole configuration on the parameter $m$, while keeping other parameters constant. Specifically, it demonstrates the existence of a minimum mass $m_{min}$ and a maximum mass $m_{max}$ where, depending on the value of $m$, the black hole can exhibit two horizons, become extremal, or transform into a naked singularity.
	
\begin{figure}[h]
		\centerline{\includegraphics[width=0.7\textwidth]{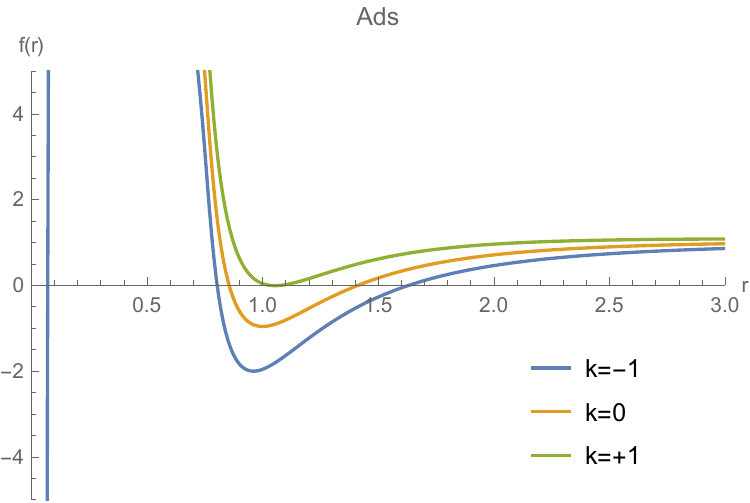}}
		\caption{$f(r)$ for asymptotically AdS spacetime versus $r$ for different values of $k$ with $q=1$, $m=2.5$, $\hat{\mu}_2=0.04$, $\hat{\mu}_3=-0.001$, $\hat{\mu}_4=-0.0002$, $\hat{\mu}_5=-0.0001$.}
		\label{fig:Adsfvsrfork}
	\end{figure}

	Furthermore, the roots of the function $f(r)$, which correspond to the locations of the horizons, vary with the parameter $k$. This variation is depicted in Figure \eqref{fig:Adsfvsrfork}, where it can be observed that for the case $k=-1$, there are three distinct horizons.
	
	This phenomenon highlights the advantage of the quintic (and also quartic \cite{Ghanaatian:2018gdl}) QTG over lower-order theories. In the case of $k=-1$, the outermost horizon represents the event horizon, while no information is available regarding the other two inner horizons. Consequently, any conclusions about the interior of the black hole remain elusive.
	
\begin{figure}
		\includegraphics[width=7.5cm, height=6cm]{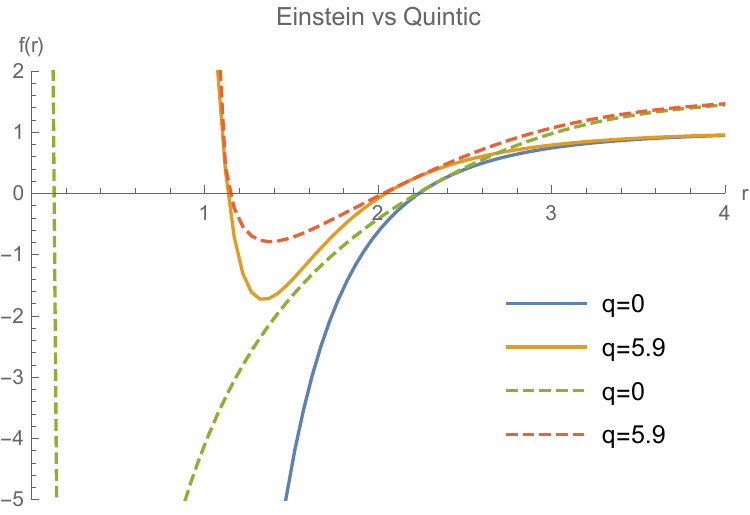}
		\vspace{-0.5cm} 
		\includegraphics[width=7.5cm, height=6cm]{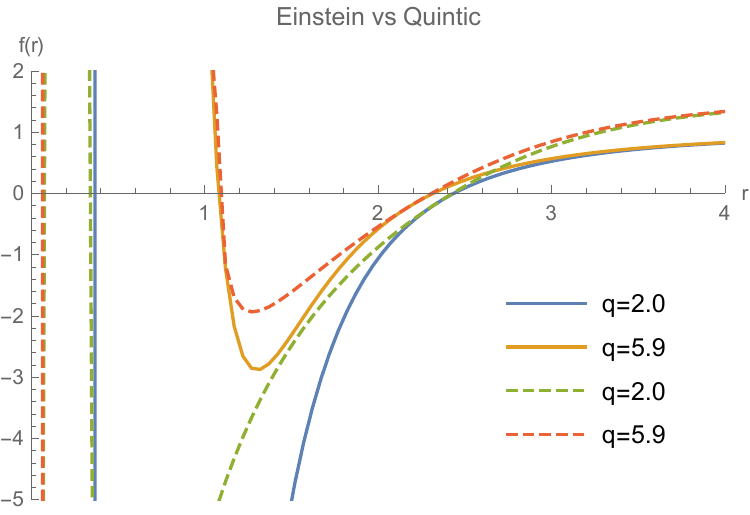}
		\caption{Plot of asymptotically AdS solution of $f(r)$ in Einstein (solid lines) and 5QT gravity (dashed lines) versus $r$ for different values of $q$ and $k=1$ (left), $k=-1$ (right), $\hat{\mu}_2=0.4$, $\hat{\mu}_3=-0.1$, $\hat{\mu}_4=-0.0002$ and $\hat{\mu}_5=-0.0001$}
		\label{fig:AdSQuinticVsEinsteinforq}
	\end{figure}

	The behavior of the function $f(r)$ in both Einstein gravity and 5QTG is compared in Figure \eqref{fig:AdSQuinticVsEinsteinforq}. In Einstein gravity, one of the solutions of the Einstein equation is the Schwarzschild solution. The Schwarzschild black hole, characterized by $q = 0$ (zero electrical charge), exhibits a single horizon. This is clearly depicted in Figure \eqref{fig:AdSQuinticVsEinsteinforq} (left) for the case of $k=+1$, where two horizons are present only when $q \neq 0$. However, in 5QTG, interestingly, two horizons are observed even when $q=0$, which distinguishes it from Einstein theory.
	This finding suggests that higher-order QTGs, such as 4- and 5QTGs, exhibit an effect similar to that of electric charge. This is in contrast to Einstein theory, where the presence of a charge is required to generate multiple horizons. 
	Additionally, Figure \eqref{fig:AdSQuinticVsEinsteinforq} (right) demonstrates that in Einstein gravity, for the case of $q=0$ and $k=-1$, a black hole displays two horizons. However, in the context of 5QTG, three horizons are observed for the same conditions. 

	\textcolor{teal}{\subsection*{\textit{Asymptotically dS spacetimes}}}
	
\begin{figure}[h]
		\centerline{\includegraphics[width=0.7\textwidth]{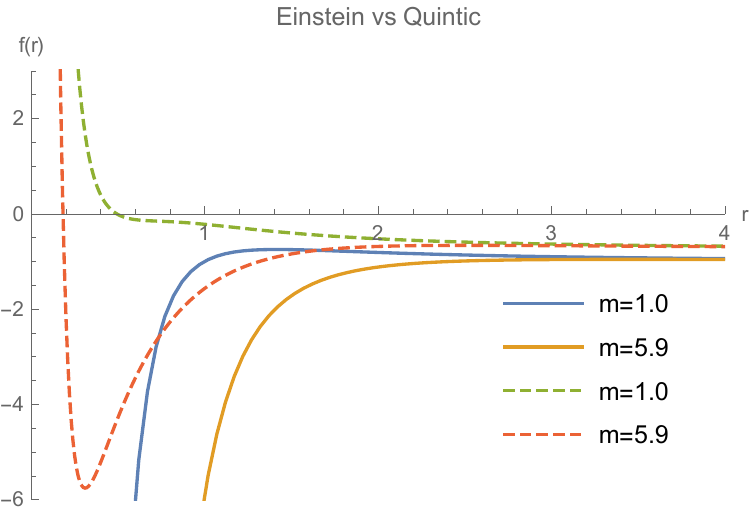}}
		\caption{Plot of asymptotically dS solution of $f(r)$ in Einstein (solid lines) and 5QT gravity (dashed lines) versus $r$ for different values of $m$ and $k=1$, $\hat{\mu}_2=0.4$, $\hat{\mu}_3=-0.1$, $\hat{\mu}_4=-0.0002$ and $\hat{\mu}_5=-0.0001$  .}
		\label{fig:dSQuinticVsEinsteinform}
	\end{figure}

	In the context of asymptotically de Sitter space, the condition $\lim_{r \rightarrow \infty} f(r) = -1$ is satisfied. Based on this, we can define the cosmological constant as follows:
	\begin{eqnarray}
		\Lambda = \frac{n(n-1)}{2L^2}(-\hat{\mu}_5 + \hat{\mu}_4 - \hat{\mu}_3 + \hat{\mu}_2 +1)~,
	\end{eqnarray}
	where, specifically for asymptotic de Sitter space with $\Lambda > 0$, we have
	\begin{eqnarray}
		-\hat{\mu}_5 + \hat{\mu}_4 - \hat{\mu}_3 + \hat{\mu}_2 > -1~.
	\end{eqnarray}
	Figure \eqref{fig:dSQuinticVsEinsteinform} demonstrates that in Einstein gravity, only a naked singularity is present, and there are no black holes. However, in the framework of 5QTG, it is possible to predict the existence of a black hole with a single horizon. This observation highlights that QTGs allow for the existence of black holes, which are forbidden in the context of Einstein gravity.
	
\begin{figure}[h]
		\centerline{\includegraphics[width=0.7\textwidth]{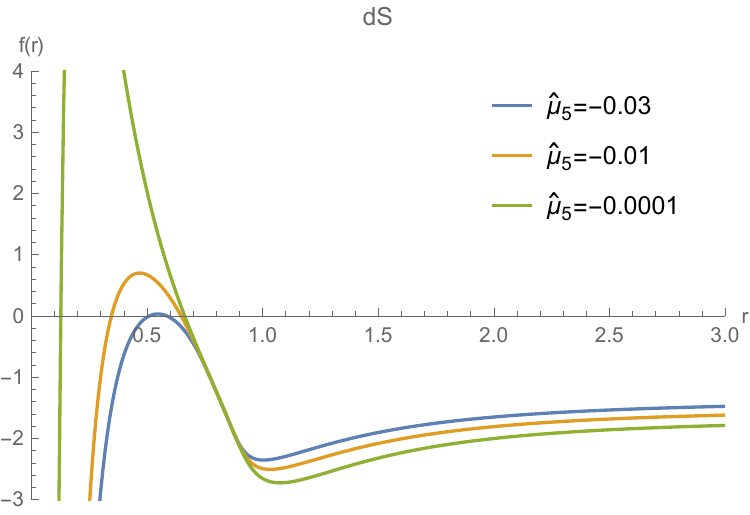}}
		\caption{Plot of asymptotically dS solution of $f(r)$ in 5QT gravity versus $r$ for different values of $\hat{\mu}_5$ for $k=-1$, $\hat{\mu}_2=-0.4$, $\hat{\mu}_3=-0.1$, and $\hat{\mu}_4=-0.002$.}
		\label{fig:dsfvsrformu5}
	\end{figure}

In order to investigate the impact of the 5QTG, we examine the behavior of the function $f(r)$ for different values of the quintic coefficient $\hat{\mu}_5$. Figure \eqref{fig:dsfvsrformu5} illustrates this behavior.
For the given values of the other parameters, it is observed that at $\hat{\mu}_5 = -0.03$, there exists an extremal black hole. When $\hat{\mu}_5$ exceeds this critical value, black holes with two horizons are found. It is noteworthy that the outer horizon remains the same for all values of $\hat{\mu}_5$, assuming the other parameters are held constant. However, the inner horizon decreases as $\hat{\mu}_5$ increases.

	\textcolor{teal}{\subsection*{\textit{Asymptotically flat spacetimes}}}
	
\begin{figure}
		\includegraphics[width=7.5cm, height=6cm]{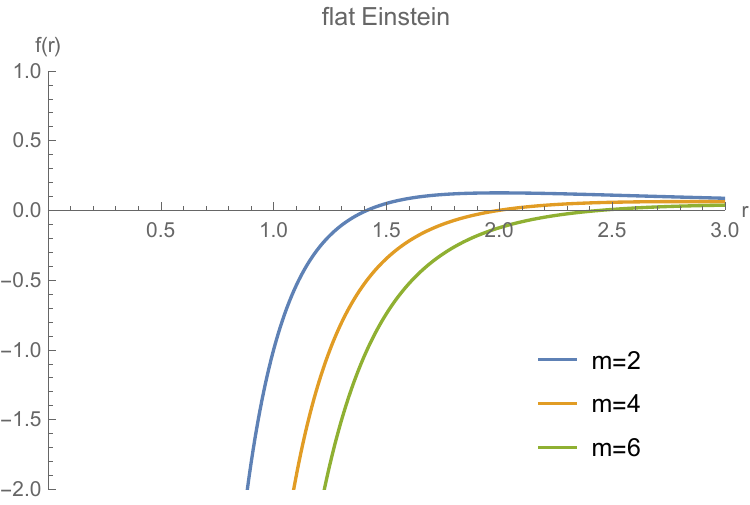}
		\vspace{-0.5cm} 
		\includegraphics[width=7.5cm, height=6cm]{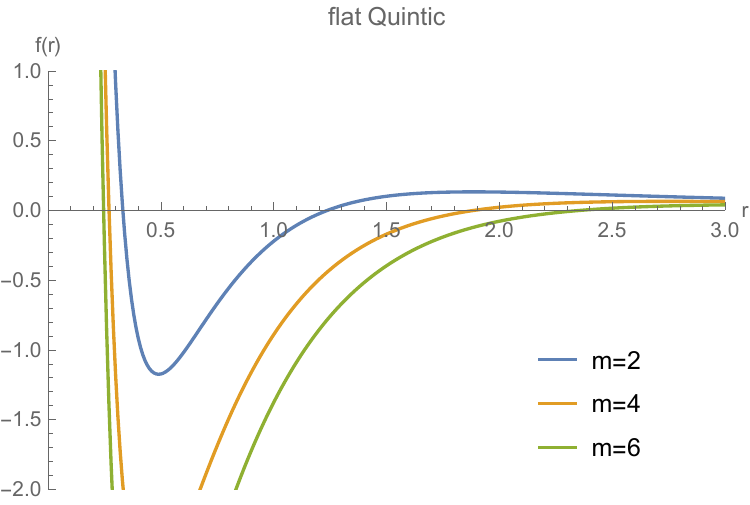}
		\caption{Plot of asymptotically flat solution of $f(r)$ in Einstein gravity (left) and 5QT gravity (right) versus $r$ for different values of $m$ and $k=1$, $\hat{\mu}_2=0.4$, $\hat{\mu}_3=-0.1$, $\hat{\mu}_4=-0.002$ and $\hat{\mu}_5=-0.001$.}
		\label{fig:flatfvsrforM}
	\end{figure}

	In Figure \eqref{fig:flatfvsrforM}, we can observe a comparison between Einstein gravity and quasi-topological gravities in the context of asymptotically flat spacetime. Asymptotically flat spacetimes are characterized by tending towards the Minkowski metric at large distances. To satisfy this condition, it is necessary to consider $\lim_{r \rightarrow \infty} f(r) = 0$, which implies a vanishing cosmological constant.
	In asymptotically flat metrics, the behavior of $f(r)$ can be approximated as $f(r) \sim L^2/r^2$, where $L$ is a characteristic length scale. By comparing the roots of $f(r)$ in Einstein gravity and 5QTG, we can observe that the roots of the latter are smaller than those of the former. This finding suggests that the size of a quintic black hole is smaller compared to an Einstein black hole.
	
	\textcolor{violet}{\section{Special Case: Analytic Solutions}\label{sec:AnalSol}}
	
	In the context of QTG, the quintic case gives rise to a fifth-order algebraic equation \eqref{equasli}, which typically requires numerical methods for solution. However, there are certain special cases where analytical solutions can be obtained. Two notable cases of interest are the pure quintic and degenerate vacuum solutions, which we will explore as follows.
	
	\textcolor{teal}{\subsection{\textit{Pure Quintic}}}
	
	In the pure 5QTG coupled to the Maxwell field, by setting $\mathcal{L}_1 = \mathcal{L}_2 = \mathcal{L}_3 = \mathcal{L}_4 = 0$, the action reduces to
	\begin{eqnarray}
		I_{\text{PM}}=\frac{1}{16\pi}\int{d^{n+1}x\sqrt{-g}\big\{-2\Lambda+\mu_{5}{\mathcal L}_5-\frac{1}{4}F^2\big\}},
	\end{eqnarray}
	where the subscript PM represents the "Pure" 5QT gravity coupled to the Maxwell field. As a result, the fifth-order algebraic equation \eqref{equasli} simplifies to 
	\begin{eqnarray}\label{5QTalg}
		\hat{\mu}_{5} \Psi^5+\kappa=0,
	\end{eqnarray}
	which can be solved analytically. This equation yields four complex roots and one real root. The real root leads to an equation for the metric function, denoted as
	\begin{eqnarray}\label{frPM}
		f_{\text{PM}}(r) = \frac{k}{r^2} + \frac{1}{\hat{\mu}_5^{1/5}}\Bigg( -\hat{\mu}_0 + \frac{m}{r^n} - \frac{q^2}{r^{2n-2}}  \Bigg)^{1/5}~,
	\end{eqnarray}
	which describes a black hole with a horizon if the equation $f(r_+) = 0$ has a real solution. The existence of a physical horizon constrains the values of the parameters $\hat{\mu}_5$, $\hat{\mu}_0$, $m$, and $q$.
	
\begin{figure}
		\includegraphics[width=7.5cm, height=6cm]{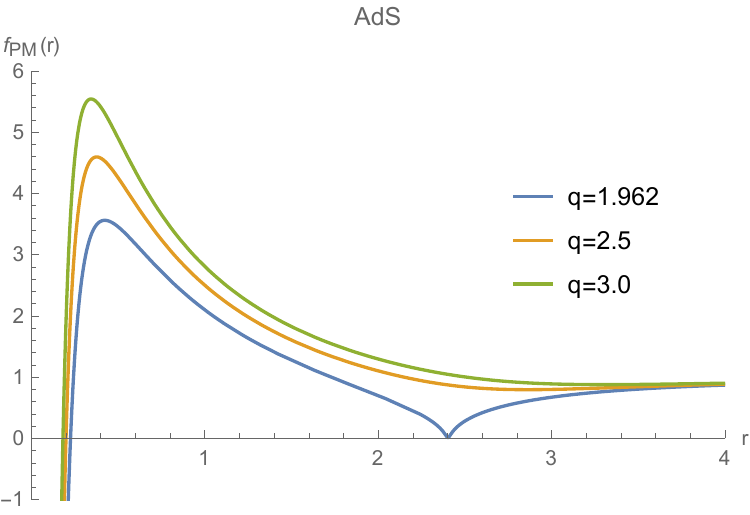}
		\vspace{-0.5cm} 
		\includegraphics[width=7.5cm, height=6cm]{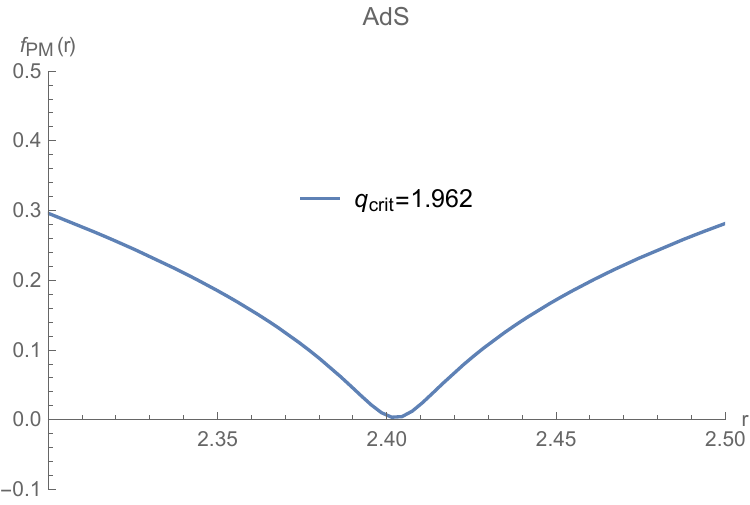}
		\caption{Plot of asymptotically AdS solution of $f_{\text{PM}}(r)$ versus $r$ in pure 5QT gravity for different values of $q$ and $k=-1$, $m=1$ and $\hat{\mu}_5=-0.01$.
			According to the left figure, it is obvious that the event horizon of the black hole decreases as $q$ increases. Also there is a critical value for the black hole charge $q_{\text{crit}} = 1.962$, for which the black hole has two horizen where the outer one is extremal which is shown in the right figure.  }
		\label{fig:PureAdS}
	\end{figure}

\begin{figure}[h]
		\centerline{\includegraphics[width=0.7\textwidth]{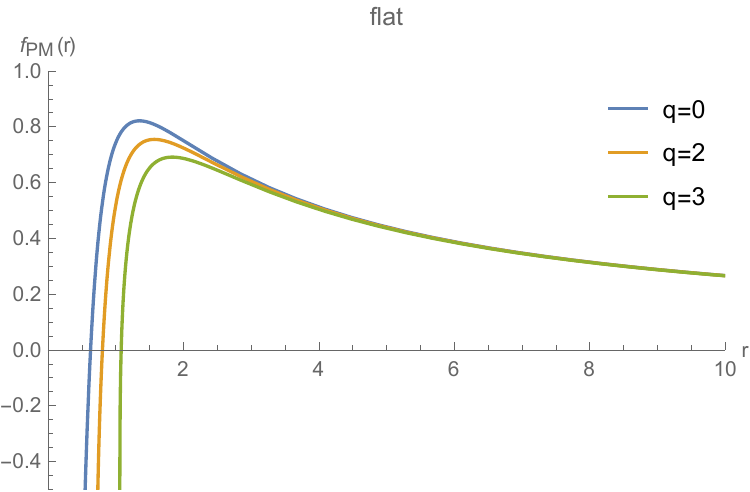}}
		\caption{Plot of asymptotically flat solution of $f_{\text{PM}}(r)$ versus $r$ in pure 5QT gravity for different values of $q$ and $k=-1$, $m=8$ and $\hat{\mu}_5=0.5$.
			It is obvious that the event horizon of the black hole increases as $q$ increases.}
		\label{fig:PureFlat}
	\end{figure}

	In the limit as $r$ approaches infinity ($r \rightarrow \infty$), the metric function takes the asymptotic form 
	\begin{eqnarray}
		f^{\infty}_{\text{PM}}(r) = \frac{k}{r^2} + \Big( - \frac{\hat{\mu}_0}{\hat{\mu}_5} \Big)^{1/5}~.
	\end{eqnarray}
	To ensure the existence of real solutions, it is necessary to impose the constraint $\hat{\mu}_5 = -\hat{\mu}_0$, which excludes de Sitter (dS) spacetime. Consequently, for the AdS and flat cases, in order for the event horizon to be physical ($r_+ > 0$), we only consider the case $k=-1$.
	Figures \eqref{fig:PureAdS} and \eqref{fig:PureFlat} depict the metric function for the pure 5QTG case in the AdS and flat spacetimes, respectively. In the AdS case, the event horizon decreases as the charge $q$ increases. On the other hand, in the flat case, the event horizon increases with increasing charge $q$.
	
	\textcolor{teal}{\subsection{\textit{Degenerate Vacuum}}}
	
	The vacuum case is another significant scenario where an analytical solution can be obtained. In this case, the black hole is characterized by the absence of both charge and mass, resulting in a pure vacuum spacetime. By considering the equation \eqref{equasli} in this context, it simplifies to
	\begin{eqnarray}\label{degvaceq}
		\hat{\mu}_{5} \Psi^5+\hat{\mu}_{4} \Psi^4+\hat{\mu}_{3} \Psi^3+\hat{\mu}_{2} \Psi^2-\Psi+\kappa_0=0,
	\end{eqnarray}
	where $\kappa_0$ represents the value of $\kappa$ from equation \eqref{kappa} when the charge and mass are set to zero:
	\begin{eqnarray}\label{kappa0}
		\kappa_0 = \kappa|_{m,q=0} = \hat{\mu}_0~.
	\end{eqnarray}
	In the subsequent analysis, we explore three distinct cases that lead to degenerate solutions in the vacuum scenario.

\textcolor{brown}{\textbf{\textit{Case a:}}}
 In the context of the vacuum case, the equation \eqref{degvaceq} exhibits a double root, resulting in two identical solutions for the metric function expressed as
 \begin{eqnarray}\label{2degroots}
 	(\Psi - \alpha)^2 (a \Psi^3 + b \Psi^2 + c\Psi + d) = 0~.
 \end{eqnarray}
 Here, $\alpha$ represents the double root, and the coefficients $a$, $b$, and $c$ are constants with $a \neq 0$.
 
\begin{figure}[h]
		\centerline{\includegraphics[width=0.7\textwidth]{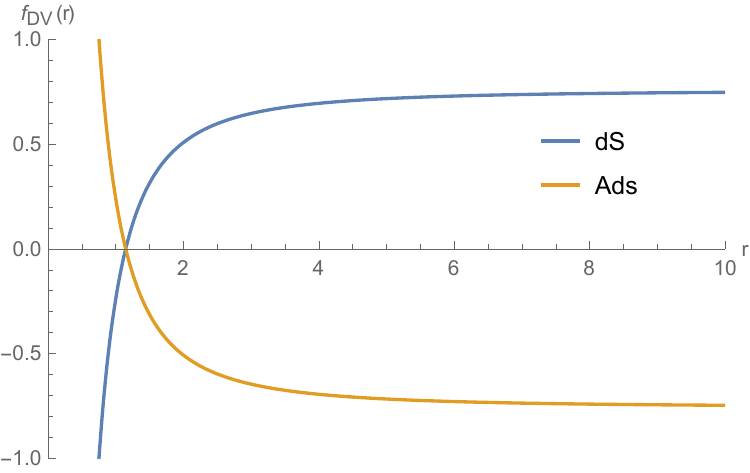}}
		\caption{Plot of the degenerate metric function in vacuum for the case for the case $\hat{\mu}_4 = \hat{\mu}_3 = \hat{\mu}_2 = 0$ for both dS and AdS spacetimes which both have the same root $r_+ = 2^{1/5} = 1.1487 $.}
		\label{fig:DV1}
	\end{figure}

 In this analysis, we focus on the scenario where all $\hat{\mu}_i$ values are zero, except for $\hat{\mu}_5$. By setting $\hat{\mu}_2 = \hat{\mu}_3 = \hat{\mu}_4 = 0$, we obtain the equation 
 \begin{eqnarray}\label{degvac1}
 	\hat{\mu}_{5} \Psi^5-\Psi+\kappa_0 \equiv (\Psi - \alpha)^2 (a \Psi^3 + b \Psi^2 + c\Psi + d) = 0~,
 \end{eqnarray}
 leading to 
 \begin{eqnarray}
 	\hat{\mu}_5 &=& \pm \hat{\mu}_0 = \frac{2^{8/5}}{5}~,\\
 	\alpha &=& \pm \frac{1}{2^{2/5}}~,
 \end{eqnarray}
 where the (+) and (-) signs correspond to de Sitter (dS) and anti-de Sitter (AdS) spacetimes, respectively.
 Consequently, we can express the metric function as 
 \begin{eqnarray}\label{fDV}
 	f_{\text{DV}}(r) = \alpha + \frac{k}{r^2}~,
 \end{eqnarray} 
 representing the Degenerate Vacuum (DV) case. Using the derived parameters, for both dS and AdS spacetimes, the root of the metric function is expressed as 
 \begin{eqnarray}\label{fDVroot}
 	r_+ = \sqrt{-\frac{k}{\alpha}}~.
 \end{eqnarray} 
 To ensure a real positive root, the signs of $\alpha$ and $k$ must be opposite. Hence, for the dS case, $k=-1$, and for the AdS case, $k=+1$. In both cases, we have 
 \begin{eqnarray}
 	r_+ = \frac{1}{\sqrt{|\alpha|}} = 2^{1/5} = 1.1487~.
 \end{eqnarray}
 Figure \eqref{fig:DV1} illustrates the plot of the degenerate metric function for both dS and AdS spacetimes.
	
	\textcolor{brown}{\textbf{\textit{Case b:}}} In this case, we focus on the scenario where there are two degenerate roots as given by equation \eqref{2degroots}, specifically considering $\hat{\mu}_3 = \hat{\mu}_4 = 0$.
	
\begin{figure}[h]
		\centerline{\includegraphics[width=0.7\textwidth]{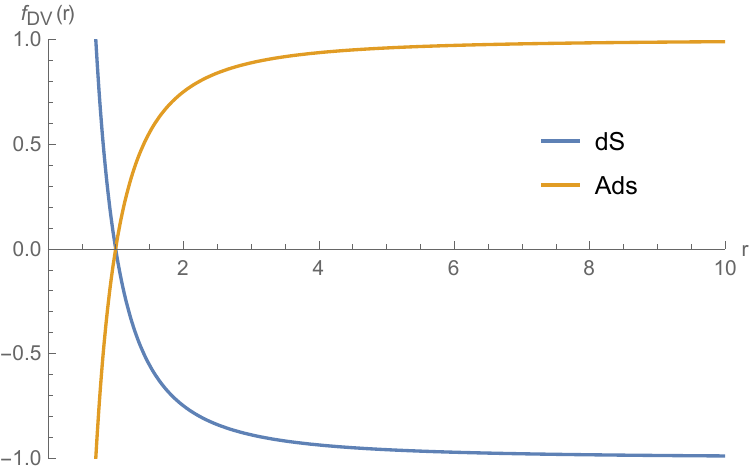}}
		\caption{Plot of the degenerate metric function in vacuum for both dS and AdS spacetimes for the case $\hat{\mu}_4 = \hat{\mu}_3 = 0$ and $\hat{\mu}_5^{\text{dS(AdS)}} =1/5 $ and $\hat{\mu}_2^{\text{dS(AdS)}} = \mp 1$ which gives $r_+ = 1$.}
		\label{fig:DV2}
	\end{figure}

	By setting $\hat{\mu}_4 = \hat{\mu}_3 = 0$, we obtain the equation 
	\begin{eqnarray}
		\hat{\mu}_{5} \Psi^5 + \hat{\mu}_{2} \Psi^2-\Psi+\kappa_0 \equiv (\Psi - \alpha)^2 (a \Psi^3 + b \Psi^2 + c\Psi + d) = 0~.
	\end{eqnarray} 
	Imposing the condition that the left-hand side of the equation has a quotient of $(\Psi - \alpha)^2$ with no remainder, we can express $\hat{\mu}_2$ and $\hat{\mu}_5$ in terms of $\alpha$ and the root of $f(r)$, denoted as $r_+ = \sqrt{-k/\alpha}$, as 
	\begin{eqnarray}
		\hat{\mu}_2 &=& \frac{4\alpha^5 \pm 1}{\alpha(3\alpha^5 \pm 2)} =-\frac{r_+^2(4k^5 \mp r_+^{10})}{k(3k^5 \mp 2r_+^{10})} ~,\\
		\hat{\mu}_5 &=& \frac{-\alpha}{3\alpha^5 \pm 2} = -\frac{k r_+^8}{3k^5 \mp 2r_+^{10}}~,\\
		\hat{\mu}_0 &=& \mp \hat{\mu}_5~,
	\end{eqnarray}
	where the upper and lower signs represent anti-de Sitter (AdS) and de Sitter (dS) spacetimes, respectively. In order for $\hat{\mu}_2$ and $\hat{\mu}_5$ to be non-singular, it can be determined that for the AdS case, $k=-1$, and for the dS case, $k=+1$. This leads to the expression 
	\begin{eqnarray}
		\hat{\mu}_2^{\text{(dS)}} = - 	\hat{\mu}_2^{\text{(AdS)}}=-\frac{r_+^2(4 + r_+^{10})}{(3 + 2r_+^{10})} ~, \\
		\hat{\mu}_5^{\text{(dS)}} = \hat{\mu}_5^{\text{(AdS)}} =-\frac{r_+^8}{3 + 2r_+^{10}}~.
	\end{eqnarray}
	For instance, considering $r_+ = 1$, the equation \eqref{equasli} becomes 
	\begin{eqnarray}
		-\frac{1}{5}\Psi^5 \pm \Psi^2 -\Psi \mp \frac{1}{5} = 0~,
	\end{eqnarray}
	where the upper and lower signs represent the AdS and dS cases, respectively. The degenerate metric function for both cases is plotted in Figure \eqref{fig:DV2}.
	
	\textcolor{brown}{\textbf{\textit{Case c:}}} In this scenario, we examine the case where all the roots of equation \eqref{degvaceq} are identical \cite{Cisterna:2017umf}, resulting in 
	\begin{eqnarray}\label{singledegeq}
		\hat{\mu}_{5} \Psi^5+\hat{\mu}_{4} \Psi^4+\hat{\mu}_{3} \Psi^3+\hat{\mu}_{2} \Psi^2-\Psi+\kappa_0 = \beta (\Psi - \alpha)^5~.
	\end{eqnarray} 
	This enables us to determine the coefficients $\hat{\mu}_i$ as shown in the following equations:
	\begin{eqnarray}
		\hat{\mu}_5 &=& -\frac{1}{5\alpha^4} = - \frac{r_+^8}{5k^4}~,\\
		\hat{\mu}_4 &=& \frac{1}{\alpha^3} = - \frac{r_+^6}{k^3}~,\\
		\hat{\mu}_3 &=& -\frac{2}{\alpha^2} = -\frac{2r_+^4}{k^2}~,\\
		\hat{\mu}_2 &=& \frac{2}{\alpha} = -\frac{2r_+^2}{k}~,\\
		\hat{\kappa_0} &=& \frac{\alpha}{5} = - \frac{k}{5r_+^2}~.
	\end{eqnarray}
	For instance, if we consider $r_+ = 1$, the corresponding equation becomes \begin{eqnarray}
		- \frac{1}{5} \Psi^5 \pm  \Psi^4 -2 \Psi^3 \pm 2 \Psi^2 - \Psi \pm \frac{1}{5} =0~,
	\end{eqnarray} 
where the (+) and (-) signs in	represent the AdS and dS spacetimes, respectively. The plot of the metric function remains the same as depicted in Figure \eqref{fig:DV2}.

	\textcolor{violet}{\section{Thermal Stability}\label{thermStab}\label{sec:Thermal}}
	In this section, we analyze the thermal stability of solutions by examining the behavior of the energy $m(S,q)$ with respect to small variations in the thermodynamic variables $S$ and $q$. To study this, we utilize the Hessian matrix in the canonical ensemble, which can be expressed as 
	\begin{eqnarray}
		H=\left[
		\begin{array}{ccc}
			H_{11} & H_{12}\\
			H_{21} & H_{22}
		\end{array} \right],
	\end{eqnarray}
	as a function of the two extensive parameters $S$ and $q$. The elements of the Hessian matrix are given by 
	\begin{eqnarray}
		H_{11}&=&\Big(\frac{\partial ^2 M}{\partial S^2}\Big)=\frac{1}{A^3 (n-1)^2 \pi}\Big[r^{(18-3n)}\sum_{i=1}^{5} A_i r^{2i} + r^{(6-n)}\sum_{i=1}^{10} B_i r^{2i} \Big] \,\,,\nonumber \\
		H_{22}&=&\Big(\frac{\partial ^2 M}{\partial Q^2}\Big)=\frac{16 \pi }{ r^{n-2}\sqrt{n^2-3 n+2}}\sqrt{\frac{n-1}{n-2}}\,\,,\,\,\nonumber \\
		H_{12}&=&H_{21}=\Big(\frac{\partial ^2 M}{\partial S\partial Q}\Big)=-\frac{64 \pi  Q }{(n-1)A r^{2n-11}}\,\,.\,\,
	\end{eqnarray} 
	Here, $A=r^8 +2 a k l^2 r^6 - 3 b k^2 l^4 r^4 + 4 c k^3 l^6 r^2 -5 e k^4 l^8$, and the coefficients $A_i$ and $B_i$ are provided in Table \eqref{tab:Hessian}.
	
	\begin{table}[!hpt]
		\begin{center}
			\begin{tabular}{|c|c|}
				\hline \textbf{Coefficient}&\textbf{Term}\\ \hline \hline
				$A_{1}$ &$-640 \pi ^2 e k^4 l^8 (2 n-11) Q^2$ \\
				\hline
				$A_{2}$ &$ 512 \pi ^2 c k^3 l^6 (2 n-9) Q^2$ \\
				\hline
				$A_{3}$  &$ -384 \pi ^2 b k^2 l^4 (2 n-7) Q^2$ \\
				\hline
				$A_{4}$ &$256 \pi ^2 a k l^2 (2 n -5) Q^2$ \\
				\hline
				$A_{5}$ &$128 \pi ^2 (2 n-3) Q^2$ \\
				\hline
				$B_{1}$ &$-5 e^2 k^9 l^{18} \left(n^2-11 n+10\right)$ \\
				\hline
				$B_{2}$ &$c e k^8 l^{16} \left(7 n^2-87 n+80\right)$\\
				\hline	
				$B_{3}$ &$ -4 k^7 l^{14} (n-1) \left[c^2 (n-8)-15 b e\right]$\\
				\hline	
				$B_{4}$ &$k^6 l^{12} (n-1) [b c (5 n-48)-a e (11 n+40)]$\\
				\hline
				$B_{5}$ &$ -k^5 l^{10} (n-1) \left\{3 b^2 (n-6)-2 [a c (n+16)-e (13 n+10)]\right\}$\\
				\hline
				$B_{6}$ &$-k^4 l^8 (n-1) [-3 a b (n-8)-c (13 n+16)+45 e \mu_0 n]$\\
				\hline	
				$B_{7}$ &$ -2 k^3 l^6 (n-1) \left\{a^2 (n-4)+2 [b (n+3)-7 c \mu_0 n]\right\}$\\
				\hline	
				$B_{8}$ &$-k^2 l^4 (n-1) [a (n-8)+15 b \mu_0 n]$\\
				\hline
				$B_{9}$ &$-k l^2 (n-1) [n (1 - 6 a \mu_0)-2]$\\
				\hline	
				$B_{10}$ &$\mu_0 (n-1) n$\\
				\hline	
			\end{tabular}
			\caption{The terms $A_{i}$ and $B_i$ of the $H_{11}$ array of the Hessian matrix.} \label{tab:Hessian}
		\end{center}
	\end{table}
	
	To determine the thermal stability of charged black holes in 5QTG, we need to consider regions where both the determinant of the Hessian, $det(H)$, and the temperature $T_+$ are positive. The positive value of $det(H)$ ensures stability, while negative temperatures are not physically meaningful.
	
\begin{figure}
		\includegraphics[width=5cm, height=4cm]{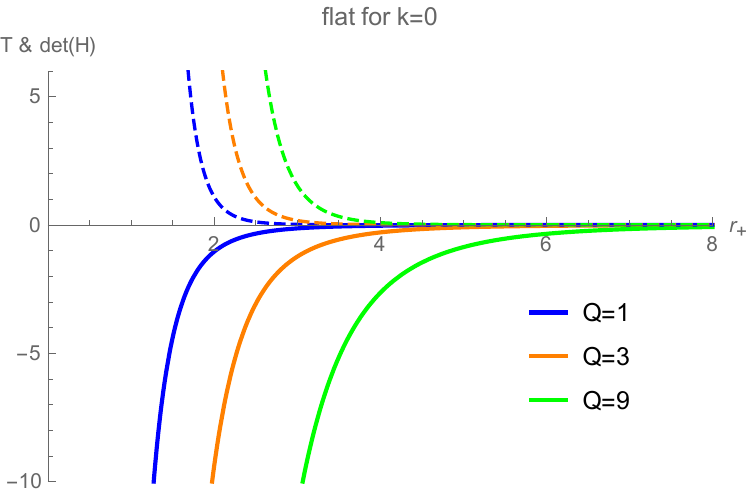}
		\includegraphics[width=5cm, height=4cm]{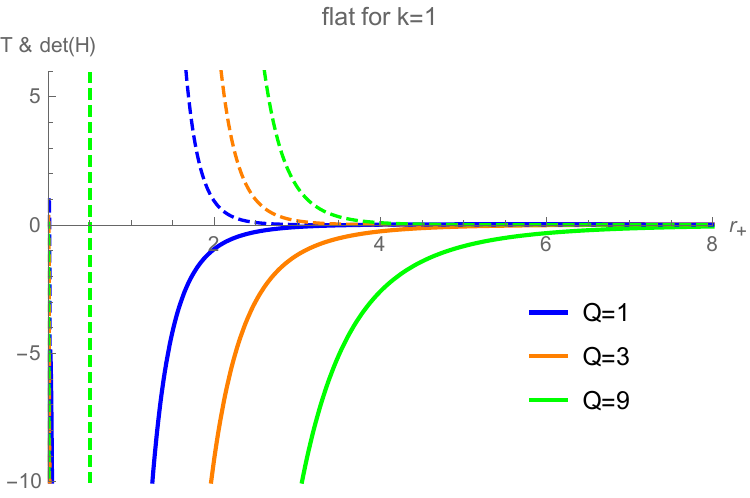}
		\includegraphics[width=5cm, height=4cm]{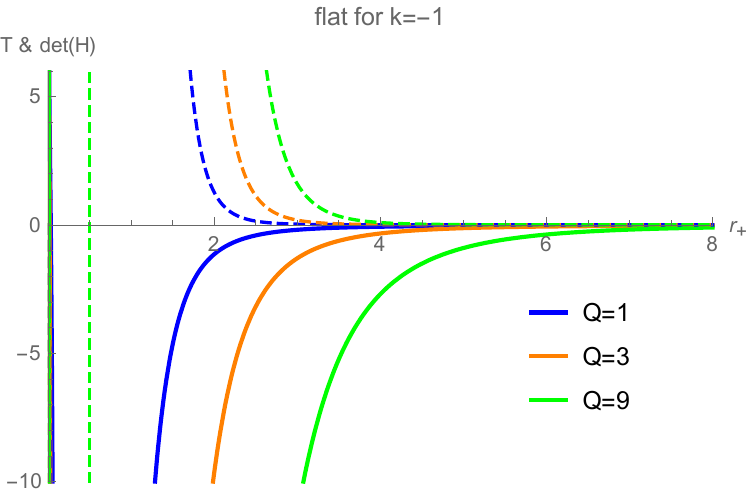}
		\includegraphics[width=5cm, height=4cm]{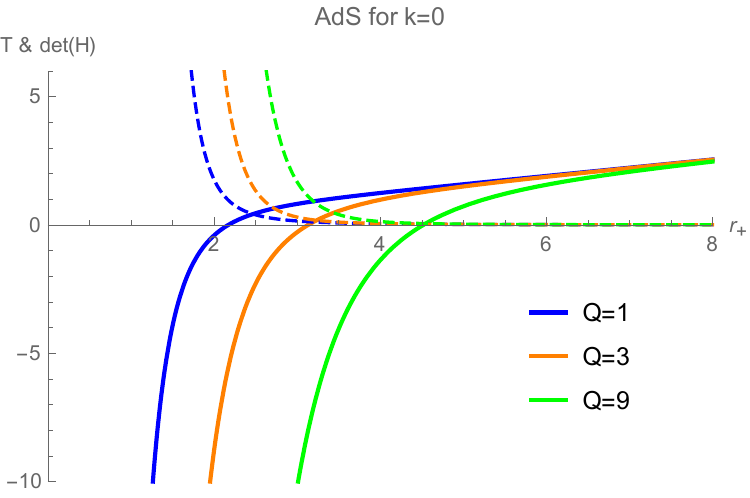}
		\includegraphics[width=5cm, height=4cm]{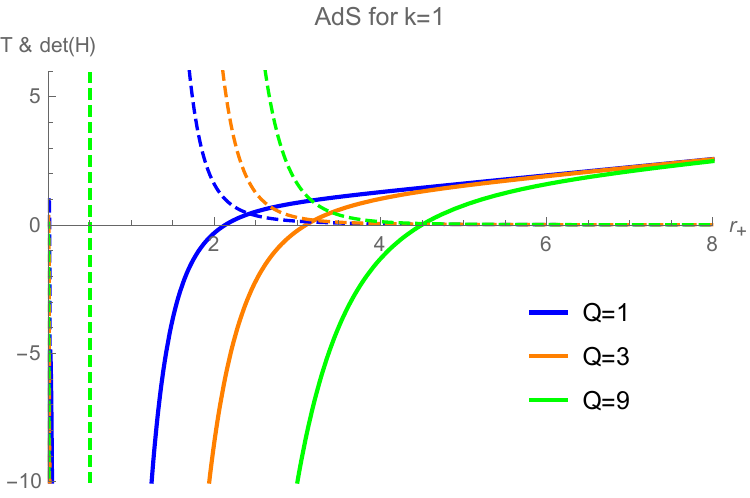}
		\includegraphics[width=5cm, height=4cm]{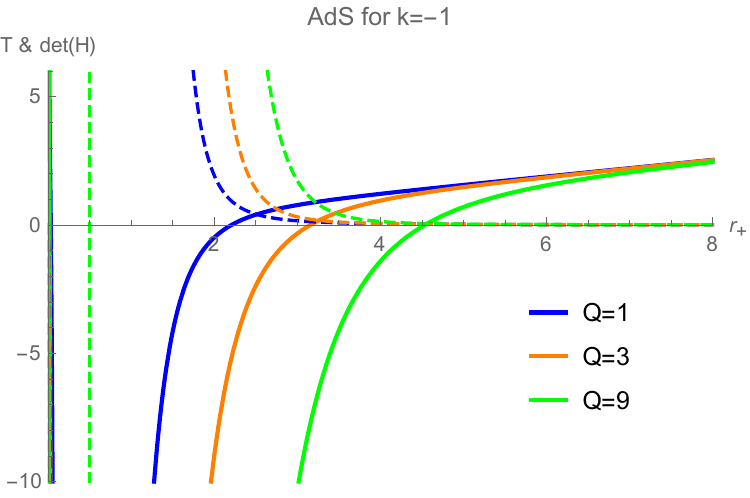}
		\includegraphics[width=5cm, height=4cm]{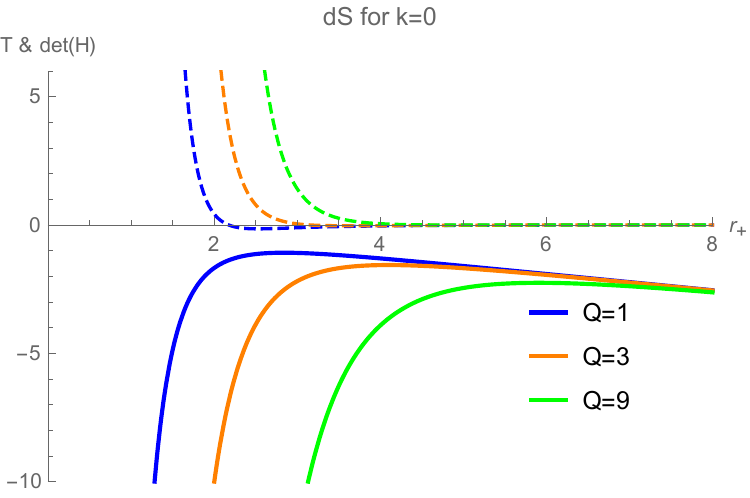}
		\includegraphics[width=5cm, height=4cm]{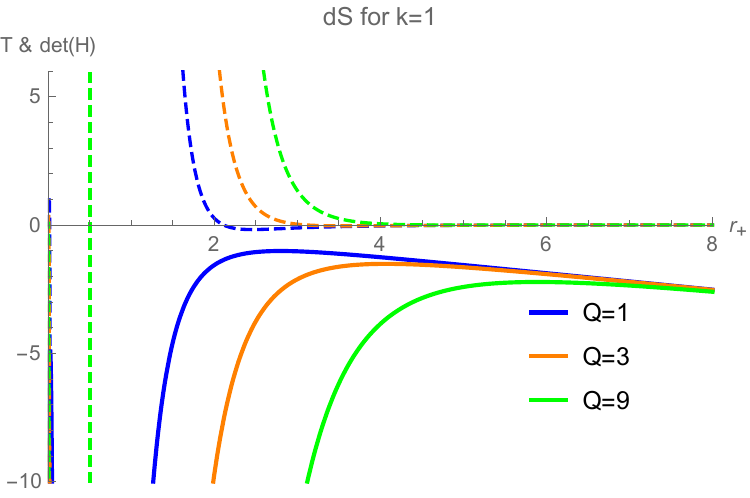}
		\includegraphics[width=5cm, height=4cm]{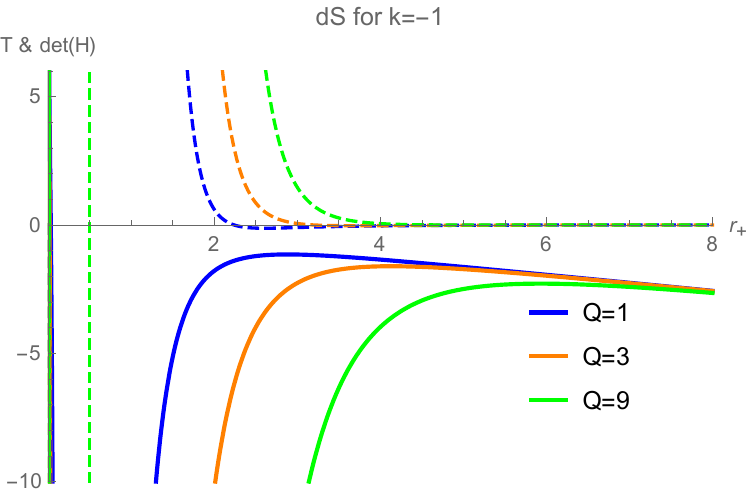}
		\caption{Plots of $T$ (thick lines) and $det(H)$ (dashed lines) vs $r_+$ for flat, AdS and dS spacetimes and $k=0, \pm1$ for different values of $Q$.}
		\label{fig:detHT}
	\end{figure}

	To investigate the stability of these black holes, we provide plots in Figure \eqref{fig:detHT} for $L=1$ in flat, AdS, and dS spacetimes. These plots illustrate the regions where both $det(H)$ and $T_+$ are simultaneously positive, indicating the thermal stability of the black hole solutions.
	
	By varying the parameter $Q$ while keeping the values of $\hat{\mu}_5$ to $\hat{\mu}_2$ fixed for flat and dS solutions with $k=\pm1$ and 0, it is observed that although the determinant of the Hessian, $det(H)$, may be positive for certain values of $r_+$, there is no region where both $T_+$ and $det(H)$ are simultaneously positive. This implies that no stable black hole solution can exist in these cases.
	
	For AdS solutions, the determinant of the Hessian, $det(H)$, is positive for most values of $r_+$, indicating potential stability. In this case, the positive value of the temperature $T_+$ defines the stability. There exists a minimum value of $r_+$, denoted as $r_{+min}$, for which $T_+$ is positive for any value of $Q$ when $r_+ > r_{+min}$. 
	
	Increasing the parameter $Q$ leads to an increase in the value of $r_{+min}$. Therefore, for small values of $Q$, there is a larger region of stability. As illustrated in Figure \eqref{fig:detHT} (middle), unlike the $k=0$ case where $det(H)$ is positive for all values of $r_+$, in the $k=-1$ case, positivity of $det(H)$ occurs only for $r_+ > r_{+1}$. The value of $r_{+1}$ is approximately the same for all values of $Q$ and is less than 2. 
	
	Regardless of the value of $k$, the temperature $T_+$ exhibits a similar behavior, with $r_{+min} > 2$ for all values of $Q$. It is noteworthy that a singularity occurs for the temperature at $r_+ = r_{+d}$, where $T_+$ goes to positive and negative infinity as $r_+$ approaches $r_{+d}$ from the left and right sides, respectively. Additionally, there is a singularity in $det(H)$ at $r_{+e}$, where $T_+ > 0$ for $r_{+e} < r_+ < r_{+max}$. Moreover, $det(H)$ becomes negative at $r_{+max}$, and it is evident that $r_{+max}$ increases with an increase in $Q$.
	
\begin{figure}
		\includegraphics[width=8cm, height=6cm]{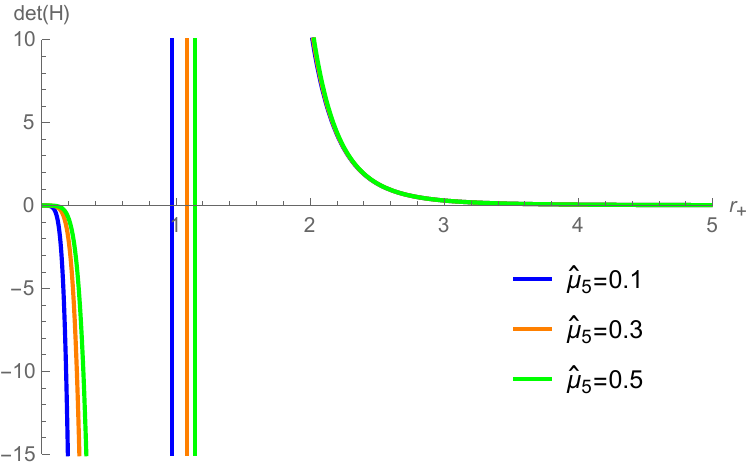}
		\includegraphics[width=8cm, height=6cm]{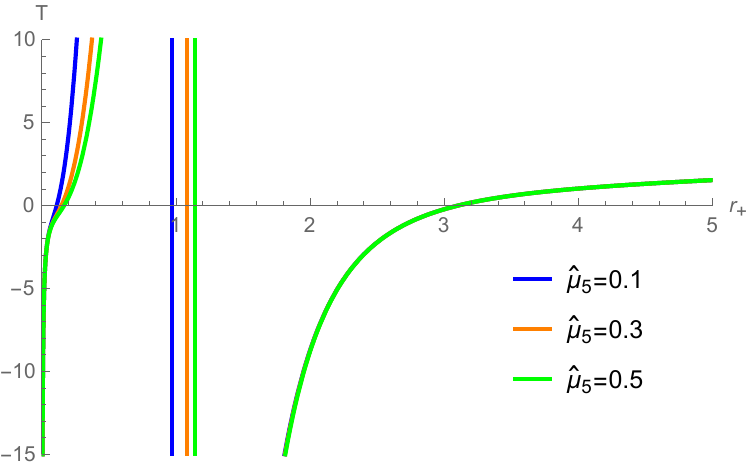}
		\caption{Plots of $det(H)$ (left) and $T$ (right) versus $r_+$ for dS solution in 5QT gravity  for different values of $\hat{\mu}_5$ for $k=1$, $\hat{\mu}_2=0.002$, $\hat{\mu}_3=-0.01$, and $\hat{\mu}_4=-0.09$.}
		\label{fig:detHTrplus}
	\end{figure}

	To explore the influence of the 5QTG parameters on the stability, we present plots in Figure \eqref{fig:detHTrplus} for different values of $\hat{\mu}_5$. These plots illustrate how the parameters affect the stability of the black hole, with the regions of positive temperature ($r_+ > r_{+min}$) indicating stability.
	For each value of $\hat{\mu}_5$, there is a singularity in $det(H)$ at $r_{+s}$. As $r_+$ approaches $r_{+s}$, $det(H)$ tends to positive infinity for $r_+ > r_{+s}$ and negative infinity for $r_+ < r_{+s}$. 
	Increasing the value of $\hat{\mu}_5$ leads to a decrease in the value of $r_{+s}$. This implies that as $\hat{\mu}_5$ increases, the singularity in $det(H)$ occurs at smaller values of $r_+$.
Additionally, there is a singularity in the temperature at $r_{+b}$, where the behavior of $T_+$ differs for $r_+ < r_{+b}$ and $r_+ > r_{+b}$. For $r_+ < r_{+b}$, $T_+$ is positive. However, for $r_+ > r_{+b}$, the positivity of $T_+$ depends on the specific value of $r_+$, with $T_+$ being positive for $r_+ > r_{+c} \approx 3.2$.

Comparing the behaviors of $det(H)$ and $T_+$ reveals that both $det(H)$ and $T_+$ are simultaneously positive for $r_+ > r_{+c} \approx 3.2$, indicating stability. This stability is independent of the value of $\hat{\mu}_5$.

	
	
	
	

	\textcolor{violet}{\section{Conclusions}\label{sec:conclusion}\label{sec:Conclusion}}
	In this study, a family of charged black hole solutions within the framework of 5QTG has been introduced. By incorporating quintic curvature terms into the action, second-order equations of motion have been derived for spacetimes with dimensions $5 \le [\frac{n}{2}]$, except for the case of $n=10$. These theories exhibit interesting properties that align with the principles of the AdS/CFT correspondence.
	
	To investigate the physical properties of these solutions, numerical analyses have been performed in three distinct spacetime backgrounds: AdS, dS, and flat spacetime. Depending on the charge parameter $q$, while keeping other parameters fixed, various scenarios arise. These scenarios include non-extremal black holes for $q < q_{min}$, black holes with two horizons for $q_{min} < q < q_{ext}$, extremal black holes for $q = q_{ext}$, or naked singularities for $q > q_{ext}$.
	
	These results provide insights into the rich and diverse behavior of the charged black hole solutions in 5QTG, allowing for a better understanding of the physical implications and implications for the AdS/CFT correspondence.
	
	In this study of Reissner-Nordström black holes in 5QTG, several noteworthy findings have been obtained. Firstly, it was discovered that in the absence of electric charge ($q=0$) and for $k=1$ (corresponding to Schwarzschild black holes), the black hole solution is not limited to a single horizon as initially expected. Instead, 5QTG has the ability to effectively mimic an electric charge, allowing for the possibility of black holes with two horizons.
	
	Comparisons were made between the behavior of 5QTG and other theories, such as 4QTG, 3QTG, and Einstein gravity. It was observed that 4- and 5QTGs possess the remarkable property of generating black holes with three horizons for $k = -1$, while 3QTG and Einstein gravity are limited to two horizons.
	
	Special cases of 5QTG were also investigated analytically, including the pure quintic and vacuum degenerate cases. In the vacuum degenerate case, which excludes mass and charge contributions, a single horizon was found, in contrast to the cases involving mass and charge.
	
	The thermal stability of the obtained solutions was explored, revealing that thermal stability is exclusively observed in AdS backgrounds, while dS and flat spacetime solutions do not exhibit thermal stability. Interestingly, solutions with smaller values of the charge parameter $q$ demonstrated larger regions of thermal stability compared to those with larger $q$. Furthermore, it was discovered that variations in the parameter $\hat{\mu}_5$ had no effect on the thermal stability. These results highlight the potential of QTG in generating intriguing solutions for charged black holes. Further investigations into the implications of this gravity theory in the presence of nonlinear electrodynamics would be worthwhile.
	
	In conclusion, this study of Reissner-Nordström black holes in 5QTG has provided valuable insights into the diverse nature of gravity beyond Einstein's theory. The emergence of charged black holes with multiple horizons, the exploration of their thermodynamic properties, and the examination of stability characteristics have advanced our understanding of the interplay between gravity, quantum field theory, and thermodynamics. Further investigations in this field hold great promise for unraveling the profound implications of these intriguing findings.
	
	\textcolor{violet}{\section*{Appendix}\label{app}}
	The coefficients $a_{i}$'s, $b_{i}$'s a nd $c_{i}$'s are defined as:
	\begin{eqnarray}
		a_{1}&=& 1,\,\,\,\,\,a_{2}=\frac{3(3n-5)}{8(2n-1)(n-3)},\,\,\,\,\,a_{3}=-\frac{3(n-1)}{(2n-1)(n-3)},\,\,\,\,\,a_{4}=\frac{3(n+1)}{(2n-1)(n-3)},\nonumber\\
		&& a_{5}=\frac{6(n-1)}{(2n-1)(n-3)},\,\,\,\,\,
		a_{6}=-\frac{3(3n-1)}{2(2n-1)(n-3)},\,\,\,\,\,a_{7}=\frac{3(n+1)}{8(2n-1)(n-3)}
	\end{eqnarray}
	\begin{table}[!hpt]
		\begin{center}
			\begin{tabular}{|c|c|c|}
				\hline \textbf{Label}&\textbf{Term}&$a_i$\\ \hline \hline
				$\mathcal{L}^{(3)}_{1}$ &$R_{ab}^{cd}R_{cd}^{ef}R_{ef}^{ab}$ &$1$ \\
				\hline
				$\mathcal{L}^{(3)}_{2}$ &$R_{abcd}R^{abcd}R$& $\frac{3(3n-5)}{8(2n-1)(n-3)}$ \\
				\hline
				$\mathcal{L}^{(3)}_{3}$  &$R_{abcd}R^{abc}{{}_e}R^{de}$& $-\frac{3(n-1)}{(2n-1)(n-3)}$ \\
				\hline
				$\mathcal{L}^{(3)}_{4}$ &$R_{abcd}R^{ac}R^{bd}$ & $\frac{3(n+1)}{(2n-1)(n-3)}$\\
				\hline
				$\mathcal{L}^{(3)}_{5}$ &$R_a{{}^b}R_b{{}^c}R_{c}{{}^a}$ &$\frac{6(n-1)}{(2n-1)(n-3)}$\\
				\hline
				$\mathcal{L}^{(3)}_{6}$ &$R_a{{}^b}R_b{{}^a}R$ &$-\frac{3(3n-1)}{2(2n-1)(n-3)}$\\
				\hline
				$\mathcal{L}^{(3)}_{7}$ &$R^3$ &$\frac{3(n+1)}{8(2n-1)(n-3)}$\\
				\hline	
			\end{tabular}
			\caption{The terms of $\mathcal{L}^{(4)}_{i}$ and their
				coefficients.} \label{tab:cubic}
		\end{center}
	\end{table}
	\pagebreak

	\begin{table}[!hpt]
		\begin{center}
			\begin{tabular}{|c|c|c|}
				\hline \textbf{Label}&\textbf{Term}&$b_i$\\ \hline \hline
				$\mathcal{L}^{(4)}_{1}$ &$R_{abcd}R^{cdef}R^{hg}{{}_{ef}}R_{hg}{{}^{ab}}$ &$-(n-1)(n^7-3n^6-29n^5+170n^4-349n^3+348n^2-180n+36)$ \\
				\hline
				$\mathcal{L}^{(4)}_{2}$ &$R_{abcd}R^{abcd}R_{ef}{{R}^{ef}}$& $-4(n-3)(2n^6-20n^5+65n^4-81n^3+13n^2+45n-18)$ \\
				\hline
				$\mathcal{L}^{(4)}_{3}$  &$RR_{ab}R^{ac}R_c{{}^b}$& $-64(n-1)(3n^2-8n+3)(n^2-3n+3)$ \\
				\hline
				$\mathcal{L}^{(4)}_{4}$ &$(R_{abcd}R^{abcd})^2$ & $-(n^8-6n^7+12n^6-22n^5+114n^4-345n^3+468n^2-270n+54) $\\
				\hline
				$\mathcal{L}^{(4)}_{5}$ &$R_{ab}R^{ac}R_{cd}R^{db}$ &$16(n-1)(10n^4-51n^3+93n^2-72n+18)$\\
				\hline
				$\mathcal{L}^{(4)}_{6}$ &$RR_{abcd}R^{ac}R^{db}$ &$-32(n-1)^2(n-3)^2(3n^2-8n+3)$\\
				\hline
				$\mathcal{L}^{(4)}_{7}$ &$R_{abcd}R^{ac}R^{be}R^d{{}_e}$ &$64(n-2)(n-1)^2(4n^3-18n^2+27n-9)$\\
				\hline
				$\mathcal{L}^{(4)}_{8}$ &$R_{abcd}R^{acef}R^b{{}_e}R^d{{}_f}$ &$-96(n-1)(n-2)(2n^4-7n^3+4n^2+6n-3)$\\
				\hline
				$\mathcal{L}^{(4)}_{9}$ &$R_{abcd}R^{ac}R_{ef}R^{bedf}$ &$16(n-1)^3(2n^4-26n^3+93n^2-117n+36)$\\
				\hline
				$\mathcal{L}^{(4)}_{10}$ &$R^4$ &$n^5-31n^4+168n^3-360n^2+330n-90$\\
				\hline
				$\mathcal{L}^{(4)}_{11}$ &$R^2 R_{abcd}R^{abcd}$ &$2(6n^6-67n^5+311n^4-742n^3+936n^2-576n+126)$\\
				\hline
				$\mathcal{L}^{(4)}_{12}$ &$R^2 R_{ab}R^{ab}$ &$8(7n^5-47n^4+121n^3-141n^2+63n-9)$\\
				\hline
				$\mathcal{L}^{(4)}_{13}$ &$R_{abcd}R^{abef}R_{ef}{{}^c{{}_g}}R^{dg}$ &$16n(n-1)(n-2)(n-3)(3n^2-8n+3)$\\
				\hline
				$\mathcal{L}^{(4)}_{14}$ &$R_{abcd}R^{aecf}R_{gehf}R^{gbhd}$ &$8(n-1)(n^7-4n^6-15n^5+122n^4-287n^3+297n^2-126n+18)$\\
				\hline
			\end{tabular}
			\caption{The terms of $\mathcal{L}^{(4)}_{i}$ and their
				coefficients.} \label{tab:quart}
		\end{center}
	\end{table}
	\pagebreak
	
	In Eq. (\eqref{quintic}), the coefficients $c_{i}$ are:
	\begin{center}
		\begin{longtable}{|c|c|c|}
			\caption{The terms of $\mathcal{L}^{(5)}_{i}$ and their
				coefficients.}\label{tab:quint}\\
			\hline
			\textbf{Lable} & \textbf{Term} &  $ c_i $ \\
			\hline\hline
			\endfirsthead
			\multicolumn{3}{c}%
			{\tablename\ \thetable\ -- \textit{Continued from previous page}} \\
			\hline
			\textbf{Lable} & \textbf{Term} & $ c_i $ \\
			\hline
			\endhead
			\hline \multicolumn{3}{r}{\textit{Continued on next page}} \\
			\endfoot
			\hline
			\endlastfoot
			$\mathcal{L}^{(5)}_{1}$ & $R R_{b}^{a} R_{c}^{b} R_{d}^{c} R_{a}^{d}$ & $ \begin{array} {lcl}
				\frac{1}{n-2}( 22\,{n}^{12}+98\,{n}^{11}-4227\,{n}^{10}+26488\,{n}^{9}-34298\,{n}^{8} \\ -314764\,{n}^{7}+1879963\,{n}^{6}-5179230\,{n}^{5}+8667296\,{n}^{4} \\ -9278000\,{n}^{3}+6209228\,{n}^{2}-2352032\,n+379200)
			\end{array} $
			\\
			\hline
			$\mathcal{L}^{(5)}_{2}$ & $ R R_{b}^{a} R_{a}^{b} R_{ef}^{cd} R_{cd}^{ef}$ & $ \begin{array} {lcl} 9\,{n}^{11}+34\,{n}^{10}-1541\,{n}^{9}+11499\,{n}^{8}-25758
				\,{n}^{7}-81964\,{n}^{6} \\ +660233\,{n}^{5} -1886059\,{n}^{4}+3046869\,{n}
				^{3}-2977682\,{n}^{2} \\ +1666312\,n-41192 \end{array} $ \\
			\hline
			$\mathcal{L}^{(5)}_{3}$  & $R R_{c}^{a} R_{d}^{b} R_{ef}^{cd} R_{ab}^{ef}$ & $ \begin{array} {lcl} \frac{1}{2(n-2)}( -58\,{n}^{12}-162\,{n}^{11}+10663\,{n}^{10}-84812\,{
					n}^{9} \\ +229322\,{n}^{8}+436556\,{n}^{7}-5176607\,{n}^{6}+18005330\,{n}^
				{5} \\ -35943244\,{n}^{4}+45563680\,{n}^{3} -36695932\,{n}^{2}+17330208\,n \\ -
				3674560) \end{array} $ \\
			\hline
			$\mathcal{L}^{(5)}_{4}$ & $R_{b}^{a} R_{a}^{b} R_{d}^{c}  R_{e}^{d} R_{c}^{e}$ & $ \begin{array} {lcl} -\frac{2(n-1)}{n-2}( 9\,{n}^{11}+34\,{n}^{10}-1541\,{n}^{9}+11499\,{n}^{8}-25758\,{n}^{7} \\ -81964\,{n}^{6} +660233
				\,{n}^{5}-1886059\,{n}^{4}+3046869\,{n}^{3} \\ -2977682\,{n}^{2}+1666312\,n  -411920) \end{array} $\\
			\hline
			$\mathcal{L}^{(5)}_{5}$ & $R_{b}^{a} R_{c}^{b} R_{a}^{c}  R_{fg}^{de} R_{de}^{fg}$ & $ \begin{array} {lcl} \frac{1}{4(n-2)}(208\,{n}^{13}-4737\,{n}^{12}+40968\,{n}^{11}-159932\,{n}^{10} \\ +101251\,{n}^{9} +1850607\,{n}^{8}-9772230\,{n}^{7}+27253898
				\,{n}^{6} \\ -50334197\,{n}^{5} +65342916\,{n}^{4}-60349728\,{n}^{3}+
				38913248\,{n}^{2} \\ -16207200\,n +3316864) \end{array} $ \\
			\hline
			$\mathcal{L}^{(5)}_{6}$ & $ R_{b}^{a} R_{d}^{b} R_{f}^{c}  R_{ag}^{de} R_{ce}^{fg}$ & $ \begin{array} {lcl} \frac{1}{n-2}(-296\,{n}^{13}+5380\,{n}^{12}-47491\,{n}^{11}+235224\,{n}^{10} \\ -501416\,{n}^{9} -1195535\,{n}^{8}+12548311\,{n}^{7}-45635482
				\,{n}^{6} \\  +100350946\,{n}^{5}  -146329207\,{n}^{4}+143219210\,{n}^{3} \\ -
				91132732\,{n}^{2}+34380784\,n-5893664) \end{array} $\\
			\hline
			$\mathcal{L}^{(5)}_{7}$ & $R_{b}^{a} R_{d}^{b} R_{f}^{c} R_{cg}^{de} R_{ae}^{fg}$ & $ \begin{array} {lcl} \frac{1}{4(n-2)}(184\,{n}^{14}-3251\,{n}^{13}+28056\,{n}^{12}-109604\,{n}^{11} \\ +6501\,{n}^{10}+1605461\,{n}^{9}-5747494\,{n}^{8}+3380818\,{
					n}^{7} \\ +36068661\,{n}^{6} -144617644\,{n}^{5}+293846956\,{n}^{4} \\ -
				373665156\,{n}^{3}+300154032\,{n}^{2} -139373056\,n \\ +28457216) \end{array} $\\
			\hline
			$\mathcal{L}^{(5)}_{8}$ & $R_{b}^{a} R_{c}^{b} R_{ae}^{cd} R_{gh}^{ef} R_{df}^{gh}$ & $ \begin{array} {lcl} \frac{1}{4(n-2)}(596\,{n}^{13}-7977\,{n}^{12}+71966\,{n}^{11}-505962\,{n}^{10} \\ +2089493\,{n}^{9} -2365377\,{n}^{8}-20061508\,{n}^{7}+
				119539756\,{n}^{6} \\ -335696499\,{n}^{5} +581268584\,{n}^{4}-651267392\,{n
				}^{3} \\ +461836336\,{n}^{2}-188752160\,n +33860352) \end{array} $\\
			\hline
			$\mathcal{L}^{(5)}_{9}$ & $R_{b}^{a} R_{c}^{b} R_{ef}^{cd} R_{gh}^{ef} R_{ad}^{gh}$ & $ \begin{array} {lcl} \frac{1}{16(n-2)}(-184\,{n}^{15}+3155\,{n}^{14}-31372\,{n}^{13}+214234\,{n}^{12} \\ -1011489\,{n}^{11}+2804783\,{n}^{10}+374252\,{n}^{9}-44192768\,{n}^{8} \\ +224431715\,{n}^{7}-655954220\,{n}^{6}+1293485398\,{n
				}^{5} \\ -1792474880\,{n}^{4} +1739485312\,{n}^{3}-1131595440\,{n}^{2} \\ +
				442875968\,n-78459392) \end{array} $\\
			\hline
			$\mathcal{L}^{(5)}_{10}$ & $R_{b}^{a} R_{c}^{b} R_{eg}^{cd} R_{ah}^{ef} R_{df}^{gh}$ & $ \begin{array} {lcl} \frac{1}{2(n-2)}(304\,{n}^{14}-5487\,{n}^{13}+51364\,{n}^{12}-296956\,{n}^{11} \\ +1020583\,{n}^{10} -1134859\,{n}^{9}-8135394\,{n}^{8}+
				52879112\,{n}^{7} \\ -168012561\,{n}^{6} +347472004\,{n}^{5}-498259688\,{n}
				^{4} \\ +497441450\,{n}^{3}-331820224\,{n}^{2} +132631584\,n \\ -23851392) \end{array} $\\
			\hline
			$\mathcal{L}^{(5)}_{11}$ & $R_{c}^{a} R_{d}^{b} R_{ab}^{cd} R_{gh}^{ef} R_{ef}^{gh}$ & $ \begin{array} {lcl} \frac{1}{4(n-2)}(-244\,{n}^{13}+4709\,{n}^{12}-34468\,{n}^{11}+95172\,{n}^{10} \\ +152097\,{n}^{9} -1923839\,{n}^{8}+6353794\,{n}^{7}-11131154
				\,{n}^{6} \\ +10232149\,{n}^{5} -1781288\,{n}^{4}-6422656\,{n}^{3}+6551632
				\,{n}^{2} \\ -2066336\,n-21504) \end{array} $\\
			\hline
			$\mathcal{L}^{(5)}_{12}$ & $R_{c}^{a} R_{d}^{b} R_{ae}^{cd} R_{gh}^{ef} R_{bf}^{gh}$ & $ \begin{array} {lcl} \frac{1}{8(n-2)}(416\,{n}^{14}-10647\,{n}^{13}+86586\,{n}^{12} \\ -223848\,{n}^{11}-764407\,{n}^{10} +6904499\,{n}^{9}-18735836\,{n}^{8} \\ +
				11482750\,{n}^{7}+69049061\,{n}^{6} -239246282\,{n}^{5} \\ +400589060\,{n}^
				{4}-410760584\,{n}^{3}+261506352\,{n}^{2} \\ -94377920\,n+14506368) \end{array} $\\
			\hline
			$\mathcal{L}^{(5)}_{13}$ & $R_{c}^{a} R_{d}^{b} R_{ef}^{cd} R_{gh}^{ef} R_{ab}^{gh}$ & $ \begin{array} {lcl} \frac{1}{16(n-2)}(-184\,{n}^{15}+4003\,{n}^{14}-34770\,{n}^{13}+206558\,{n}^{12} \\ -1209685\,{n}^{11}+6001605\,{n}^{10}-16647870\,{n}^{9} \\ 
				-3841080\,{n}^{8}+218943659\,{n}^{7}-902806270\,{n}^{6} \\ +2083343490\,{n
				}^{5}-3136302944\,{n}^{4}+3171015856\,{n}^{3} \\ -2094129968\,{n}^{2}+
				819673024\,n-144243200) \end{array}$\\
			\hline
			$\mathcal{L}^{(5)}_{14}$ & $R_{c}^{a} R_{d}^{b} R_{eg}^{cd} R_{ah}^{ef} R_{bf}^{gh}$ & $ \begin{array} {lcl} \frac{1}{2(n-2)}(388\,{n}^{14}-4716\,{n}^{13}+29243\,{n}^{12}-136746\,{n}^{11} \\ +450540\,{n}^{10}-132929\,{n}^{9}-8134503\,{n}^{8}+48332850
				\,{n}^{7} \\ -155977854\,{n}^{6}+329810835\,{n}^{5}-478872342\,{n}^{4} \\ +
				476176930\,{n}^{3}-310557920\,{n}^{2}+119520320\,n-20516736) \end{array}$\\
			\hline
			$\mathcal{L}^{(5)}_{15}$ & $R_{c}^{a} R_{e}^{b} R_{af}^{cd} R_{gh}^{ef} R_{bd}^{gh}$ & $ \begin{array} {lcl} \frac{1}{8(n-2)}(664\,{n}^{14}-9139\,{n}^{13}+57128\,{n}^{12}-185108\,{n}^{11} \\ +159381\,{n}^{10}+1223517\,{n}^{9}-6100234\,{n}^{8}+14465638
				\,{n}^{7} \\ -20079091\,{n}^{6}+16176660\,{n}^{5}-10250920\,{n}^{4} \\ +
				14943120\,{n}^{3}-20032144\,{n}^{2}+11852416\,n-2285952) \end{array}$\\
			\hline
			$\mathcal{L}^{(5)}_{16}$ & $R_{b}^{a} R_{ad}^{bc} R_{fh}^{de} R_{ci}^{fg} R_{eg}^{hi}$ & $ \begin{array} {lcl} \frac{1}{2}(540\,{n}^{13}-10293\,{n}^{12}+81315\,{n}^{11}-291890\,{n}^{10} \\ +61415\,
				{n}^{9}+3807202\,{n}^{8}-16976001\,{n}^{7}+38237858\,{n}^{6} \\ -48573651
				\,{n}^{5}+26466351\,{n}^{4}+15817758\,{n}^{3}-37469132\,{n}^{2} \\ +
				25084592\,n-6246112) \end{array}$\\
			\hline
			$\mathcal{L}^{(5)}_{17}$ & $R_{b}^{a} R_{de}^{bc} R_{cf}^{de} R_{hi}^{fg} R_{ag}^{hi}$ & $ \begin{array} {lcl} \frac{1}{16(n-2)}(432\,{n}^{15}-4127\,{n}^{14}+19469\,{n}^{13}-170554\,{n}^{12} \\ +1675605\,{n}^{11}-9507738\,{n}^{10}+29711901\,{n}^{9}-
				40384306\,{n}^{8} \\ -54483045\,{n}^{7}+383370077\,{n}^{6}-912069066\,{n}^
				{5} \\ +1331612328\,{n}^{4}-1311989568\,{n}^{3}+870354912\,{n}^{2} \\ -
				355966464\,n+67907584) \end{array} $\\
			\hline
			$\mathcal{L}^{(5)}_{18}$ & $R_{b}^{a} R_{df}^{bc} R_{ac}^{de} R_{hi}^{fg} R_{eg}^{hi}$ & $ \begin{array} {lcl} \frac{1}{2(n-2)}(62\,{n}^{15}-261\,{n}^{14}+82\,{n}^{13}-34985\,{n}^{12}+465930\,{n}^{11} \\ -2557591\,{n}^{10}+6958394\,{n}^{9}-5370935\,{n}^{8}-27811996\,{n}^{7} \\ +116102040\,{n}^{6}-231876220\,{n}^{5}+291631996
				\,{n}^{4} \\ -242759516\,{n}^{3}+131704680\,{n}^{2}-43058976\,n+6606272) \end{array} $\\
			\hline
			$\mathcal{L}^{(5)}_{19}$ & $ R_{b}^{a} R_{df}^{bc} R_{ah}^{de} R_{ei}^{fg} R_{cg}^{hi}$ & $ \begin{array} {lcl} \frac{1}{4(n-2)}(-656\,{n}^{15}+8832\,{n}^{14}-54341\,{n}^{13}+172912\,{n}^{12} \\ -46160\,{n}^{11}-2326941\,{n}^{10}+11290819\,{n}^{9}-
				23788482\,{n}^{8} \\ +3603422\,{n}^{7}+115748491\,{n}^{6}-336672132\,{n}^{
					5} \\ +513849092\,{n}^{4}-485503192\,{n}^{3}+287262016\,{n}^{2} \\ -98692160\,
				n+15146752) \end{array}$\\
			\hline
			$\mathcal{L}^{(5)}_{20}$ & $R_{b}^{a} R_{df}^{bc} R_{gh}^{de} R_{ei}^{fg} R_{ac}^{hi}$ & $ \begin{array} {lcl} \frac{1}{8(n-2)}(-1328\,{n}^{15}+24603\,{n}^{14}-201582\,{n}^{13}+847816\,{n}^{12} \\ -1334949\,{n}^{11}-4313683\,{n}^{10}+32443416\,{n}^{9} \\ -106895622\,{n}^{8}+248652911\,{n}^{7}-456679514\,{n}^{6} \\ +663949044\,{
					n}^{5}-732715856\,{n}^{4}+587021544\,{n}^{3} \\ -326614080\,{n}^{2}+
				116003840\,n-20182272) \end{array}$\\
			\hline
			$\mathcal{L}^{(5)}_{21}$ & $R_{cd}^{ab} R_{eg}^{cd} R_{ai}^{ef} R_{fj}^{gh}R_{bh}^{ij}$ & $ \begin{array} {lcl} \frac{1}{8(n-2)}(184\,{n}^{16}-3339\,{n}^{15}+27760\,{n}^{14}-131212\,{n}^{13} \\ +355561\,{n}^{12}-357883\,{n}^{11}-1572146\,{n}^{10}+
				11274022\,{n}^{9} \\ -44976839\,{n}^{8}+128367156\,{n}^{7}-256851408\,{n}^
				{6} \\ +335947624\,{n}^{5}-240332008\,{n}^{4}+11641104\,{n}^{3} \\ +134199392
				\,{n}^{2}-104249344\,n+26702336) \end{array} $\\
			\hline
			$\mathcal{L}^{(5)}_{22}$ & $R_{ce}^{ab} R_{af}^{cd} R_{gi}^{ef} R_{bj}^{gh}R_{dh}^{ij}$ & $ \begin{array} {lcl} \frac{1}{2(n-2)}(-284\,{n}^{15}+4973\,{n}^{14}-37942\,{n}^{13}+144773\,{n}^{12} \\ -109479\,{n}^{11}-1875825\,{n}^{10}
				+12234317\,{n}^{9}-
				45166705\,{n}^{8} \\ +119677671\,{n}^{7}-240530864\,{n}^{6}+367236029\,{n}
				^{5} \\ -416310288\,{n}^{4}+337180200\,{n}^{3}-183807888\,{n}^{2} \\ +60487840
				\,n-9120896) \end{array}$\\
			\hline
			$\mathcal{L}^{(5)}_{23}$ & $R_{ce}^{ab} R_{ag}^{cd} R_{bi}^{ef} R_{fj}^{gh}R_{dh}^{ij}$ & $ \begin{array} {lcl} \frac{1}{4(n-2)}(-8\,{n}^{15}+2019\,{n}^{14}-38926\,{n}^{13}+337600\,{n}^{12} \\ -1605181\,{n}^{11}+3785705\,{n}^{10}+1659444\,{n}^{9} \\ -
				45775086\,{n}^{8}+176674471\,{n}^{7}-400284742\,{n}^{6} \\ +611871600\,{n}
				^{5}-648934536\,{n}^{4}+469233704\,{n}^{3} \\ -218274992\,{n}^{2}+57719936\,n-6343424) \end{array}$\\
			\hline
			$\mathcal{L}^{(5)}_{24}$ &$R_{ce}^{ab} R_{fg}^{cd} R_{hi}^{ef} R_{aj}^{gh}R_{bd}^{ij}$ &$ \begin{array} {lcl} \frac{1}{4}(184\,{n}^{15}-3179\,{n}^{14}+25777\,{n}^{13}-115454\,{n}^{12} \\ +228481\,
				{n}^{11}+522238\,{n}^{10}-5783003\,{n}^{9}+23848974\,{n}^{8} \\ -64717433
				\,{n}^{7}+128477225\,{n}^{6}-193789406\,{n}^{5} \\ +224224860\,{n}^{4}-
				195140632\,{n}^{3}+120313912\,{n}^{2} \\ -46440128\,n+8345216) \end{array}$\\
			\hline
		\end{longtable}
	\end{center}

	\onecolumngrid


	\onecolumngrid

\end{document}